\documentclass[twocolumn,secnumarabic,amssymb, nobibnotes, aps, prd]{revtex4-2}

\usepackage{graphicx}
\usepackage{dcolumn}
\usepackage{bm}%
\usepackage{amsmath}
\usepackage[caption = false]{subfig}
\usepackage[utf8x]{inputenc}
\usepackage{xcolor}
\usepackage{amssymb}
\usepackage{natbib}
\begin{document}

\preprint{APS/123-QED}

\title{Topological Shell Structures in Neutron Stars: Effects on Equilibrium, Oscillations, and Gravitational-Wave Signatures}

\author{Debojoti Kuzur}
 \email{dkuzur.phys@raghunathpurcollege.ac.in}
\affiliation{Department of Physics, Raghunathpur College, Purulia - 723133, West Bengal, India}%
\date{\today}
\author{Kamal Krishna Nath}
\affiliation{School of Physical Sciences, National Institute of Science Education and Research,
An OCC of Homi Bhabha National Institute, Jatni-752050, India}%
\date{\today}

\begin{abstract}
We study the structural and dynamical consequences of introducing a 
distributional density profile inside a neutron star, representing a massless, topological shell located at an arbitrary radius. We incorporate this effect into the 
structure of neutron star and construct equilibrium 
sequence for several realistic equations of state. Radial stability is examined through the Sturm-Liouville formulation of the $\ell=0$ perturbation equation, supplemented with a jump condition and imprinting distinct features on the fundamental $f$-mode spectrum. We find strong, non-monotonic variations in the mode frequency relative to 
standard no-shell models. Using first-principles scaling relations, we estimate various gravitational wave observables such as the damping time, quality factor, luminosity and characteristic strain. These observables are then compared with the sensitivity of Advanced LIGO, and third-generation detectors such as the Einstein Telescope and Cosmic Explorer. Our results demonstrate that internal topological shells can leave potentially observable signatures in the oscillation and gravitational wave properties of neutron stars.
\end{abstract}

\maketitle

\section{Introduction}
 Neutron stars (NS) are highly compact objects with layered interiors, and understanding their internal stratification is crucial for stability and astrophysical phenomena. Various studies have shown that the inner crust plays a key role in pulsar glitches and nucleosynthesis in mergers, and quantal shell effects in dense matter can strongly influence its composition \cite{Marmorini2024,PhysRevResearch.4.013026,10.1093/mnras/stae2087,PhysRevLett.133.241201,Combi_2023}. Mondal et al. (2020) \cite{2020PhRvC.102a5802M} shows that the quantal effects like shell correction or pairing may play a relevant role in determining the composition of the inner crust of the NS.

The thin-shell formalism, also known as the Israel junction method \cite{1966NCimB..44....1I,PhysRevD.72.044016,universe8050250}, offers a framework for assembling composite stellar configurations by joining two spacetime regions across a singular hypersurface (the shell), which possesses its own surface energy density and pressure. In this description, the shell is modeled as an infinitesimally thin layer represented mathematically by a delta function, which carries only a limited set of collective degrees of freedom and acts as the boundary between distinct interior and exterior geometries. Building on this idea, Zloshchastiev (1999) \cite{1999IJMPD...8..549Z} investigated spherically symmetric thin shells composed of barotropic fluid as fully general-relativistic models for NSs and circumstellar structures. Bondi (1964) \cite{1964RSPSA.282..303B} was among the first to propose a simplified stellar model in which an inner core is joined to an exterior region by a thin shell. However, a complete description of the shell’s physical properties was not provided. Later advances clarified these constructions using Israel’s junction conditions. Pereira and Rueda (2023) \cite{2023Univ....9..305P} extended the method to slowly rotating configurations, matching two neutral, asymptotically flat rotating spacetimes across a dynamical thin shell (accurate to second order in rotation) and analyzing the resulting stability and surface degrees of freedom.

Shell-based models in astrophysics have been explored in a wide range of contexts, revealing how thin layers and discontinuities can significantly influence the structure of relativistic systems \cite{10.1046/j.1365-8711.2003.06057.x,Piro_2005,Verma_2025,Gorda_2023,https://doi.org/10.1002/asna.70017,PhysRevC.107.025801,Prasad_2018}. While Mondal’s study \cite{2020PhRvC.102a5802M} relies on a semiclassical treatment of dense matter, it clearly emphasizes that shell-like features within nuclear matter must be incorporated into realistic NS models. Beyond stellar interiors, thin-shell techniques have also been applied to more exotic geometries. For instance, Eiroa et al. (2019) investigated cylindrical thin shells surrounding charged black strings in a spacetime with negative cosmological constant, examining both standard interior–exterior matchings and wormhole configurations. Likewise, Hosotani et al. (2002) \cite{2002PhRvD..66j4020H} uncovered spherical cosmic-shell solutions arising in scalar-field theories with double-well potentials; when the potential has two minima, gravity permits a thin spherical shell that transitions between a true and a false vacuum in an asymptotically de Sitter background. More recently, Arrechea et al. (2025) \cite{2025PhRvD.111l4017A} introduced bilayered stellar configurations to investigate the attainable limits of compactness. Their framework models an NS as an inner core enveloped by a denser outer layer, with the two regions sometimes joined through an infinitesimally thin shell. This construction provides a controlled setting for examining deviations from the classical assumptions underlying Buchdahl’s compactness bound.

Building upon these foundations, the present work employs the thin-shell formalism along with numerical modeling to investigate how analogous shell structures inside NS cores affect stability and oscillation spectra. The insights from these earlier studies form the basis for our approach and serve as benchmarks for understanding the dynamical role that thin shells may play within compact astrophysical objects. The study of NS stability via radial oscillations remains one of the most direct ways to probe the internal structure of compact stars. The eigenfrequencies of radial modes, particularly the fundamental $f$-mode, are sensitive to the equation of state (EoS) and internal discontinuities. In this work, we analyze NS models with a sharp pressure discontinuity introduced through a massless shell. Such discontinuities may arise from phase transitions or structural rearrangements in the dense interior. Unlike models with delta-function mass shells, our configuration maintains a continuous mass profile $M(r)$ while exhibiting a jump in pressure. This setup leads to a discontinuity in the derivative of the radial perturbation variable and introduces matching conditions essential for computing the correct eigenfrequencies.

We outline below the sequence of analyses carried out in the subsequent sections. The article is organized to move from the construction of the shell-modified equilibrium to the resulting dynamical and gravitational-wave consequences. In Section II, we formulate the density-distribution model for a massless topological shell and derive the associated pressure-jump condition required for its incorporation into the Tolman–Oppenheimer–Volkoff (TOV) equations, followed by its numerical implementation. In Section III, we present the radial stability analysis, where the Sturm–Liouville eigenvalue problem is solved together with the jump condition for the Lagrangian displacement. The resulting equilibrium sequences and their impact on the mass–radius relation and compactness are examined in Section IV. In Section V, we extend the discussion to gravitational-wave analysis, including frequency shifts, damping behavior and luminosity estimates. Finally , in section VI, we summarize the principal results and outline possible extensions.

\section{NS with a Massless Topological Pressure Shell}
We have outlined the astrophysical motivation for introducing localized internal structures and now look into the mathematical framework required to incorporate such features into relativistic stellar models. Any realistic phase boundary or topological surface layer within an NS is expected to be microscopically thin compared to the stellar radius; it is natural to describe it using the thin-shell approximation. Thus, in this section, we construct the density distribution model for a thin local shell structure inside the NS, without resolving its thickness. The gravitational Poisson equation can be written as
\begin{equation}
    \nabla^2\Phi(r)=\frac{1}{r^2}\frac{d}{dr}\left(r^2\frac{d\Phi}{dr}\right)=4\pi G\rho(r)
    \label{pois}
\end{equation}
where $\Phi$ and $\rho$ are the gravitational potential and the density profile of the NS, respectively. Although, inherently the NS interior is inherently relativistic in nature, however Poisson equation is able to capture and illustrate the imprint of such singular structures on the gravitational potential $\Phi$ through the mass density profile $\rho(r)$. For a  standard mass shell, with a profile of the form $\rho(r)=\frac{M}{4\pi r^2}\delta(r-R)$, the solution is the trivial gravitational potential
$\Phi(r)= (-\frac{GM}{r} \;{\rm for }\; r>R,  -\frac{GM}{R} \;{\rm for}\; r\le R) $, where $M$ is the gravitational mass of the NS.
However, another possible non-trivial solution to the Poisson equation that is obtained by assuming at first a solution of the form \cite{10.1093/mnras/stae1258,PhysRevD.43.1129}
\begin{equation}
    \Phi(r)=\begin{cases} 
      \Phi_+(r) & r>R_s \\
      \Phi_-(r) & r\le R_s.
   \end{cases}\end{equation}
This partitions the NS into two distinct topological regions, $\Omega_-=\{r<R_s\}$ and $\Omega_+=\{r>R_s\}$. We assume a continuity at $r=R_s$, but then allow a discontinuity in the derivative $\frac{d\Phi}{dr}\Big|_{r=R_s}$. A choice for the solution respecting the discontinuity can be written as:
\begin{equation}
    \Phi(r)=-\frac{GM}{r}\Theta(r-R_s)-\frac{GM}{R_s}\Theta(R_s-r)
\end{equation}
where $\Theta$ is the Heaviside step function. This ansatz represents a general distributional profile for a massless yet dynamically active shell. Substituting this in the Poisson equation, we get,
\begin{equation}
\frac{d}{dr}\left(r^2\frac{d\Phi}{dr}\right)  =GM\left[(1-2r)\delta(r-R_s)-r^2\delta'(r-R_s)\right] 
\end{equation}
It can be seen that the solution necessarily contains both delta-like and step-like structures, reflecting the presence of surface contributions of a shell defect. Moreover, the defect is non-removable under any smooth coordinate transformation. Upon rearrangement and comparing to equation \ref{pois}, we obtain a mass density of the form,
\begin{equation}
    \rho(r)\propto\frac{1}{r^2}\delta(r-R_s)+\frac{1}{r}\delta'(r-R_s)
\end{equation}
This forms a codimension-1 defect, embedded in a 3+1 spacetime. In order to now understand how such a shell structure affects the interior of the NS, we start with a system where the NS has hadronic matter for $0\le r <R_s$, that is, the region $\Omega_-$ and $R_s<r\le R$, which is region $\Omega_+$, where $R$ is the surface of the NS. The shell has no finite radial thickness, and we consider the NS model with such mass density given by \cite{10.1093/mnras/stae1258} (placed at $R_s$)
\begin{equation}
    \rho(r)=A\frac{\delta(R_s-r)}{r^2}+B\frac{\delta'(R_s-r)}{r}
    \label{eq1}
\end{equation}
Although $\delta(r-R_s)$ and $\delta'(r-R_s)$ in principle diverge at the shell location, $r=R_s$, they may not be interpreted as infinite physical densities, in the sense that usual distributional treatments of thin shells generally encode finite surface mass and finite jumps in derivative quantities once integrated. Thus, the shells and, following that, the NS's physical content remains completely regular despite the distributional appearance of the density, as will be seen in the following discussions. The mass enclosed by the shell could be calculated by integrating the density over the volume of the NS
\begin{equation}
    M_{enc}=\int_0^R 4\pi r^2\rho(r) dr
    \label{eq2}
\end{equation}
Using the density given in \ref{eq1}, we get two integral terms, the first one is $I_1=\int_0^R4\pi r^2A\frac{\delta(R_s-r)}{r^2}dr=4\pi A$ and the second integral is $I_2=\int_0^R 4\pi r^2B\frac{\delta'(R_s-r)}{r}dr$.
Using the standard identity $\int_{-\infty}^{\infty}\delta'(x)\phi(x)dx=-\int_{-\infty}^{\infty}\delta(x)\phi'(x)dx=-\phi'(0)$
the second integral gives $I_2=-4\pi B$, , where both $I_1$ and $I_2$ are the conserved topological charges associated with the defect. As a result of which, the mass enclosed,
\begin{equation}
    M_{enc}=I_1+I_2=4\pi(A-B)
    \label{eq7}
\end{equation}
The enclosed mass identically vanishes for $A=B$, due to the cancellation between the opposite winding number or the charges. The delta term behaves as a spherical membrane with localised energy density and has the usual effect of being gravitationally attractive.
On the other hand, the delta prime term resembles a dipole-type distribution of mass energy, having a structure of a gradient and not just localised mass. Physically, it is like a symmetric pair of opposite point masses $m$ located at $+\epsilon$ and $-\epsilon$, mass density of which is given by,
\begin{equation}
    \rho(r)=\lim_{\epsilon\rightarrow 0}\frac{m}{\epsilon}[\delta(r+\epsilon)-\delta(r-\epsilon)]=2m\delta'(r)
    \label{eq8}
\end{equation}
Such kind of a structure has the inverse effect on the mass, essentially cancelling out the attractive nature due to the first term, thus resulting in a configuration that doesn't gravitate. Such shell defects will not contribute to the total gravitational mass of the NS; however, they will have an effect on the internal structure.
We will now investigate how it affects the standard TOV equations \cite{PhysRev.55.374,PhysRev.55.364}. The pressure gradient across the NS (in $G=c=1$) is given by
\begin{equation}
    \frac{dP}{dr}=\frac{\rho(r)+P(r)}{r^2}(m(r)+4\pi r^3P(r))\left(1-\frac{2m(r)}{r}\right)^{-1}
\end{equation}
Integrating the equation \ref{eq8} across the shell using equation \ref{eq1} from $R_s-\epsilon$ to $R_s+\epsilon$
\begin{align}
    &\int_{R_s-\epsilon}^{R_s+\epsilon}\frac{dP}{dr}dr \\
    &=\int_{R_s-\epsilon}^{R_s+\epsilon}\frac{\rho(r)+P(r)}{r^2}(m(r)+4\pi r^3P(r))\left(1-\frac{2m(r)}{r}\right)^{-1}
\end{align}
where $0<\epsilon<<1$. Assuming that the change in $P$ across the shell is much less than the change in density. Also the variation of the term $\left(1-\frac{2m(r)}{r}\right)^{-1}$ is very slow across the shell, we get
\begin{equation}
    P(R_s^+)-P(R_s^-)=-\left[\frac{A}{r^4}+\frac{B}{r^2}\frac{d}{dr}\right]\left(m(r)+4\pi r^3P(r)\right)\Big|_{r=R_s}
    \label{pjump}
\end{equation}
This pressure difference shows how the delta type defect redistributes matter inside the NS by effectively generating a downward shift in the pressure profile. This acts as a  junction relation which links both sides of the pressure shell to the TOV equation. In the next section, we numerically implement this jump condition to the stellar model and investigate the oscillation spectra and stability of the NS.

\begin{figure*}
    \centering
    \includegraphics[scale=0.45]{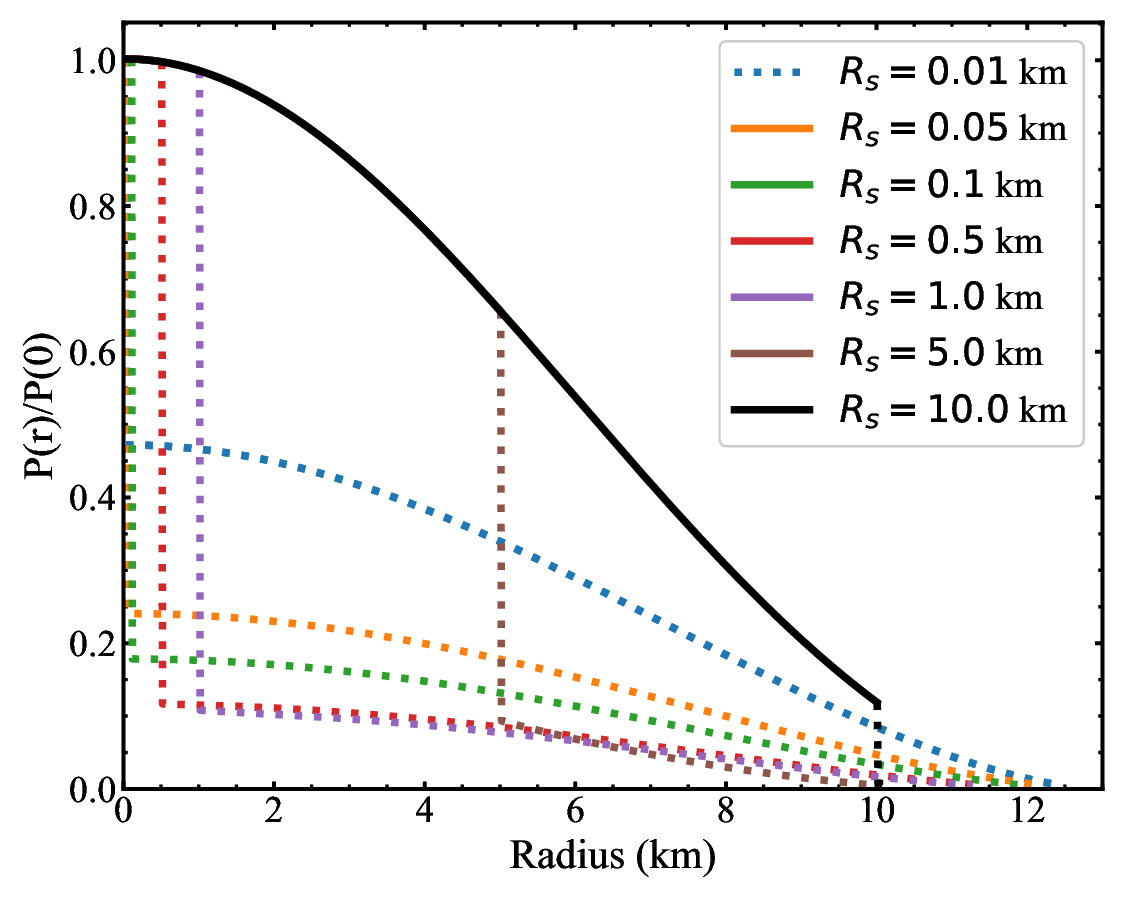}
    \includegraphics[scale=0.45]{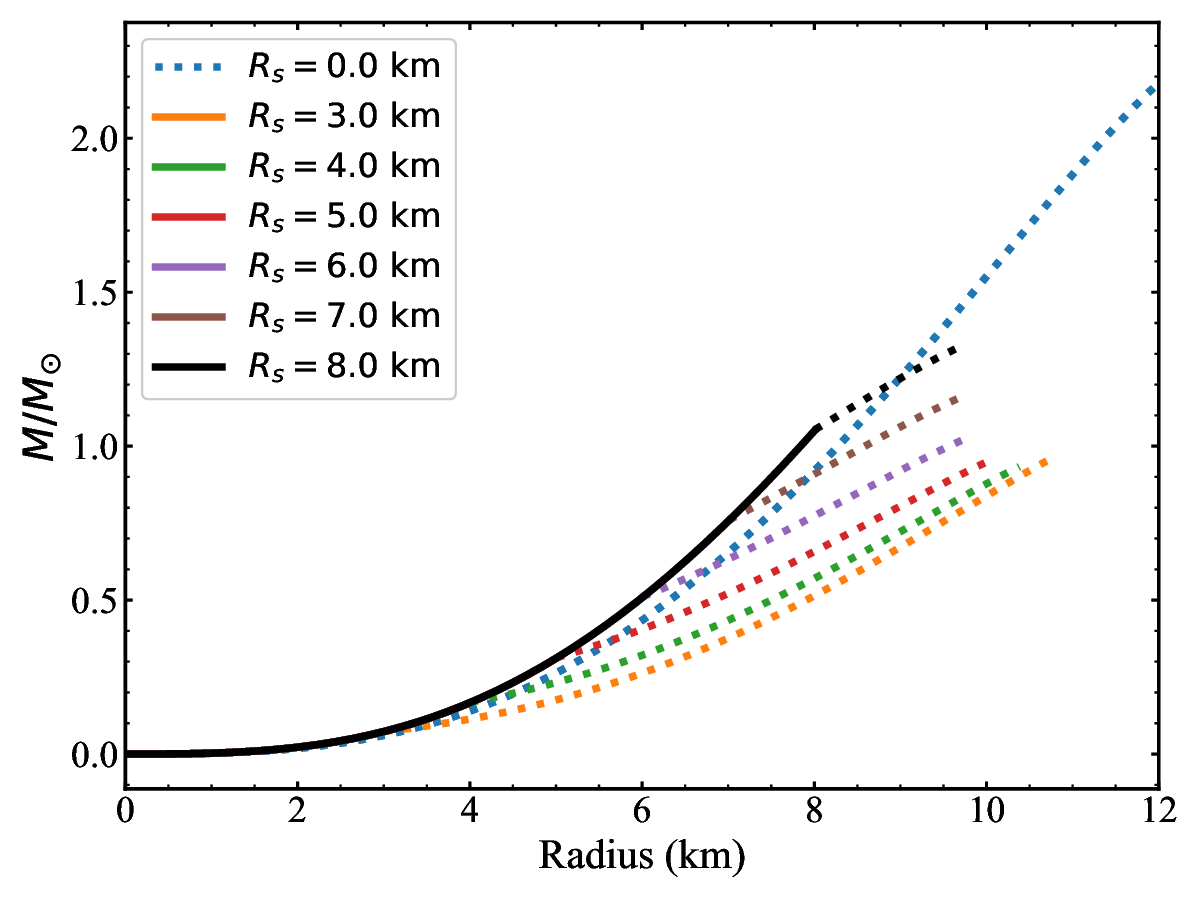}
    \caption{Left panel: Normalised pressure profiles $P(r)/P(0)$ as a function of radial coordinate for NS models with singular shells located at different shell radii $R_s$. The discontinuities in the profiles correspond to the imposed pressure jumps at the shell locations. Smaller $R_s$ lead to central suppression of pressure, whereas larger $R_s$ shift the discontinuity outward. 
Right panel: Enclosed gravitational mass $M(r)/M_\odot$ as a function of radius for the same configurations. The growth of mass halts momentarily at the shell radius due to the mass contribution of the singular layer, followed by continued accumulation in the outer layers. Models without shells ($R_s=0$) are shown as the reference case. The solid lines represent the $\Omega_-$ region whereas the dotted line represents $\Omega_+$ region.
}
\label{fig2}
\end{figure*}
\subsection{Implementation in Numerical Structure}
In order to solve the TOV equations numerically, the input data includes radius, $P(r)$, $\rho(r)$, $\phi(r)$, and $\lambda(r)$ coupled with a tabulated EoS extrapolated with cubic splines. The EoS used in this work are various field-theoretical-based relativistic mean field models, having density-dependent meson-nucleon vertex function, effective mean field models, etc, named BSR \cite{PhysRevC.76.045801}, DDME \cite{PhysRevC.66.064302}, IOPB \cite{PhysRevC.97.045806}, S27 \cite{PhysRevC.66.055803} and APR \cite{PhysRevC.58.1804}. In order to define the crust, at low densities, we have added the BPS EoS \cite{1971ApJ...170..299B}.

The Runge-Kutta method is used for numerical integration, respecting the jump in $P$. The integration takes places in three sections, first, it starts from the stellar center and then is solved outwards till the inside of the shell $R_s^-$ (region $\Omega_-$). Here standard boundary conditions are used where $M(r_{cen})=\frac{4}{3}\pi r_{cen}^3\rho(r_{cen})$ and a finite central pressure $P(r_{cen})$. Next, we come to the boundary matching at $R_s$ by calculating $P(R_s^+)$ from the equation \ref{pjump}. Here, due to the pressure jump, all the other parameters such as the metric potentials $\lambda(r)$, $\phi(r)$, and the mass $M(r)$ automatically incorporates the jump as all of them a functions of pressure. The pressure jump and the mass change can be seen in the left and right panel of figure \ref{fig2}, respectively. After the shell, from $R^+_s$ to the stellar surface $R$ (region $\Omega_+$), the integration continues with the new pressure profile till the surface is reached, defined by $P(R)\rightarrow 0$. The left panel, which represents the normalized pressure profiles for different shell radii $R_s$, displays that a small $R_s$ causes the pressure to fall much early thus reaching a zero pressure much earlier than an NS with no shell defect. This causes the NS to have a small radius and lower mass even for the same central density, while a larger radius shift the discontinuity outward into the envelope, causing the pressure discontinuity in an already low pressure region, thus not effecting the radius and mass of the NS as much. 

The right panel displays the enclosed mass $M(r)$, which remains continuous across the shell but displays a sharp change in the mass as it adds on radius by radius at $R_s$. This is a reflection of the pressure drop at $R_s$, causing the NS to acquire lower mass than a non-shell version. Next, we look into the influence of such NSs with shell on radial oscillations and discuss about the stability of such NSs.

\section{Radial Stability and Perturbation Analysis}
\subsection{Perturbation Equation}
The jump condition for pressure (equation \ref{pjump}) encapsulates the singular shell effect on the static background structure. To understand how these structural changes affect the dynamical response of the configuration, we now turn to the radial perturbation analysis, which will play a central role in determining the modified eigenfrequencies and gravitational wave (GW) properties of the NS.
For the stability analysis, we first look into the response of such NS to small radial oscillations around the equilibrium position. The motion of the perturbed fluid of the NS due to a radial (r-dependent) Lagrangian displacement can be written as \cite{1964ApJ...140..417C,Rather_2024}:  
\begin{equation}
    \Delta r(t,r)=\zeta(r)e^{i\omega t}
\end{equation}
where $\zeta(r)$ is the function which incorporates the radial profile of the oscillation and $\omega$ is the corresponding frequency mode. If the value of the frequency squared is $\omega^2 > 0$, then the oscillation corresponds to a stable oscillation, whereas, for $\omega^2<0$, the oscillation becomes unstable and leads to a stable or unstable NS-shell configuration, respectively. Assuming an adiabatic perturbation, the variation in density is related to the variation in pressure as \cite{PhysRevD.103.103003}:
\begin{equation}
    \Delta P=\gamma \frac{P}{\rho+P}\Delta\rho
\end{equation}
where $\gamma$ is the adiabatic index. This, along with the conservation of energy momentum tensor equation $\nabla_\mu T^{\mu\nu}=0$ and Einstein's Field equation in the background, we obtain the second order ODE for $\zeta(r)$
\begin{equation}
    \frac{d}{dr}\left(\Pi(r)\frac{d\zeta}{dr}\right)+\left(Q(r)+\omega^2W(r)\right)\zeta=0
\end{equation}
which has the structure of the Sturm–Liouville (SL) eigenvalue problem for $\zeta(r)$ and eigenvalue $\omega^2$ \cite{1966ApJ...145..505B,10.1093/mnras/stab050}. The functions, $\Pi$, $Q$ and $W$ depend on the background equilibrium configuration of the NS as, $\Pi(r)=\frac{\gamma Pe^{3\lambda+\phi}}{r^2}$, $Q(r)=\left[\frac{dP}{dr}\left(\frac{4}{r}+\frac{d\phi}{dr}\right)-\frac{8P}{r^2}\right]e^{3\lambda+\phi}$ and $W(r)=(\rho+P)e^{3\lambda+\phi}r^{-2}$. The unperturbed metric functions $\phi(r)$ and $\lambda(r)$ are obtained by solving the background TOV equations. Substituting the metric potentials and the functions into the Strum-Liouville eigenvalue equation, we obtain the final form of the perturbation equation as
\begin{align}
&\frac{d}{dr} \left( \gamma \frac{P}{\rho + P} r^2 e^{\phi - 3\lambda} \frac{d\zeta}{dr} \right) \nonumber \\
&+ \left[ \left( \frac{dP}{dr} \cdot \frac{d\phi}{dr} - \frac{4}{r} \frac{dP}{dr} \right) e^{\phi - 3\lambda} - \omega^2 (\rho + P) r^2 e^{\phi + \lambda} \right] \zeta = 0
\end{align}
In order to solve this numerically, we will require the jump condition of $\frac{d\zeta}{dr}$ due to the presence of the shell at $R_s$. Integrating the radial perturbation equation across an infinitesimal region containing the shell $r\in(R_s-\epsilon,R_s+\epsilon)$ and then taking $\epsilon\rightarrow 0$, will isolate the discontinuity in $\zeta'(r)$. Thus integrating gives
\begin{align}
&\left[ \frac{d\zeta}{dr} \right]_{r=R_s} \nonumber \\
&= \frac{\omega^2 e^{\lambda(R_s)}}{\gamma P(R_s) R_s^2} \left(\frac{A}{e^{\phi(R_s)}}\zeta(R_s) + B \frac{d}{dr} \left(\frac{ r\zeta(r)}{e^{\phi(r)}}\right)\Bigg|_{r=R_s} \right)
\label{strumjump}
\end{align}
We would like to find values of $\omega^2$ for which the radial perturbation equation has non-trivial solutions that satisfy proper boundary conditions and jump conditions at the shell and at the surface of the NS. For a given NS, we apply first the radial perturbation equation assuming the boundary conditions $\zeta(r)\sim r \text{ at the center and }$ $\Delta P=0 \text{ at the surface}$. The relation between the Eulerian and Lagrangian perturbations is $ \Delta P(r) = \delta P(r) + \zeta(r) \frac{dP}{dr}$ \cite{10.1111/j.1365-2966.2007.11625.x}, and the expression for the Eulerian pressure perturbation is $\delta P(r) = -\gamma P(r) \left( \frac{\zeta(r)}{r^2} \right)' - \zeta(r) \frac{dp}{dr}.$ Substituting this into the relation for $\Delta P(r)$ yields:
\begin{equation}
\Delta P(r) = -\gamma P(r) \left( \frac{\zeta(r)}{r^2} \right)'.
\end{equation} Thus, demanding $\left( \frac{\zeta(r)}{r^2} \right)'=0 $ at the surface is sufficient. We then integrate the SL differential equation from the center to just before the shell (using trial values of $\omega^2$ ) and then integrate from just after the shell to the surface, using the jump condition to link the two sides. 
\subsection{Radial Eigenfunction and Eigenvalues}
\begin{figure}
    \centering
    \includegraphics[scale=0.45]{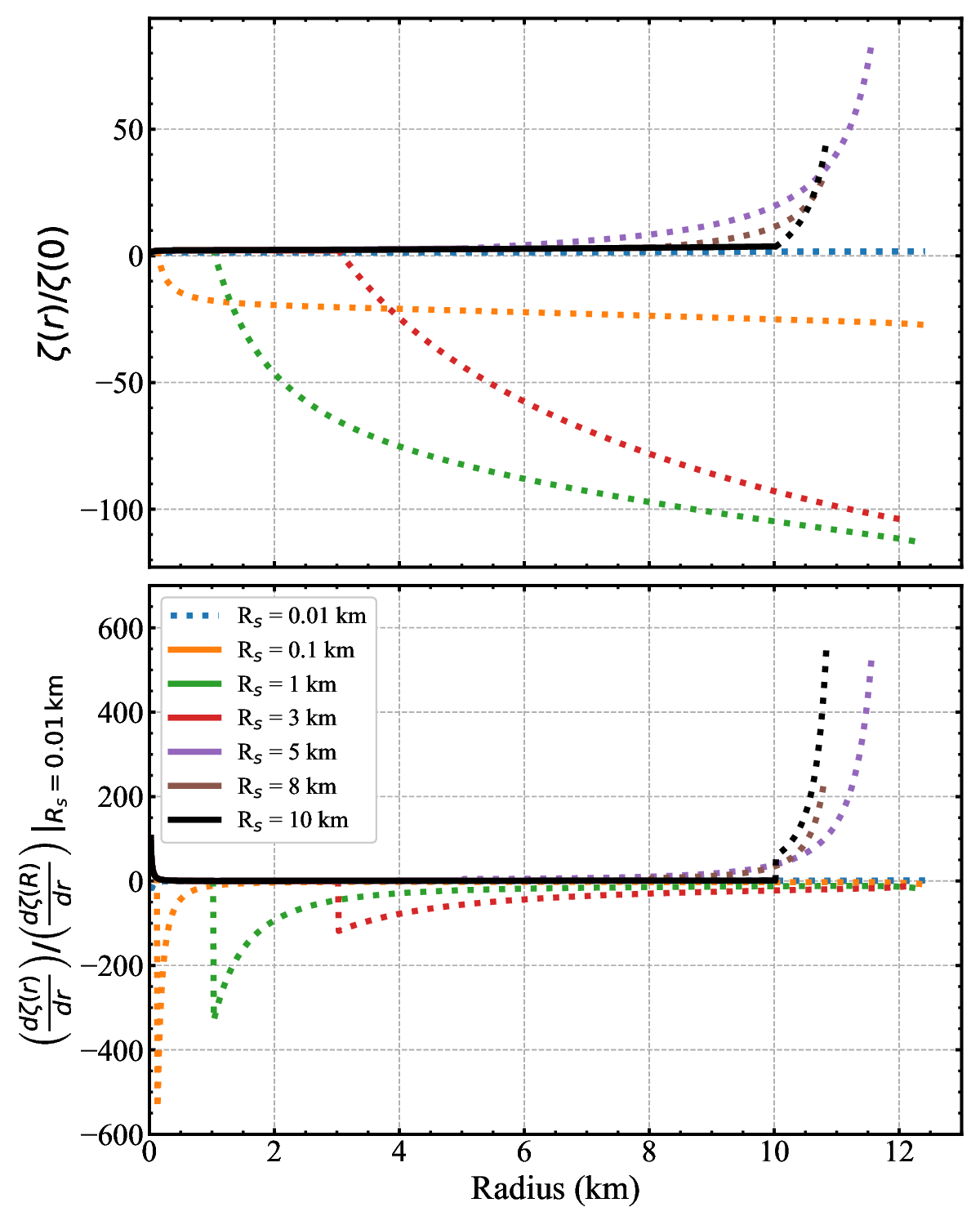}
    \caption{Radial profiles of perturbation displacement eigenfunction and its derivative for different shell radii. Top panel: Normalized radial eigenfunction $\zeta(r)/\zeta(0)$ for radial oscillations is shown for NS models containing singular mass shells located at varying shell radii $R_s$. Distinct behaviors are evident depending on the shell radius: for shells located deeper within the NS (small $R_s$), the displacement profile is largely suppressed or negative in the outer regions, whereas for outer shells (large $R_s$), the eigenfunction grows rapidly near the stellar surface, indicating stronger surface oscillatory displacement. Bottom panel: The corresponding derivative $\frac{d\zeta(r)}{dr}$ normalized with respect to its value for $R_s=0.01$ km, highlighting the effect of the shell-induced discontinuity in the derivative at the shell radius. Jumps in the derivative reflect the presence of the singular shell, introducing sharp variations in the radial displacement profile around the shell location. The solid lines represent the $\Omega_-$ region whereas the dotted line represents $\Omega_+$ region.}
    \label{fig3}
\end{figure}

By solving the SL ODE simultaneously with the TOV, the effect of the shell on the radial eigenfunction is encapsulated in figure \ref{fig3}. The figure \ref{fig3} (top panel) shows the normalized displacement $\zeta(r)/\zeta(0)$ for different shell radii $R_s$. For all the cases of different values of shell radius $R_s$, the eigenfunction remains continuous, but the overall profile gets altered 
depending on the position of the shell. When the shell lies deep in the stellar interior, the displacement in the outer layers is reduced, whereas shells placed closer to the surface enhance the amplitude of oscillations in the envelope. This illustrates how the shell redistributes oscillatory energy between the core and the outer regions.  
The figure \ref{fig3} (lower panel) represents the normalized derivative 
$(d\zeta(r)/dr)/(d\zeta(R)/dr)|_{R_s=0.01}$, which causes the effect of the jump condition (equation \ref{strumjump}) to be seen more prominently. The plot reveals a kink at the shell radius $R_s$, corresponding to the discontinuity in $\zeta'(r)$ where the magnitude of this discontinuity depends on the shell position in the density profile. Together, the two panels confirm that while the displacement itself is smooth, the shell enforces a sharp change in its slope; however, both $\zeta$ and $\zeta'$ remain finite as it reaches the surface of the NS.
\begin{figure}
    \centering
    \includegraphics[scale=0.45]{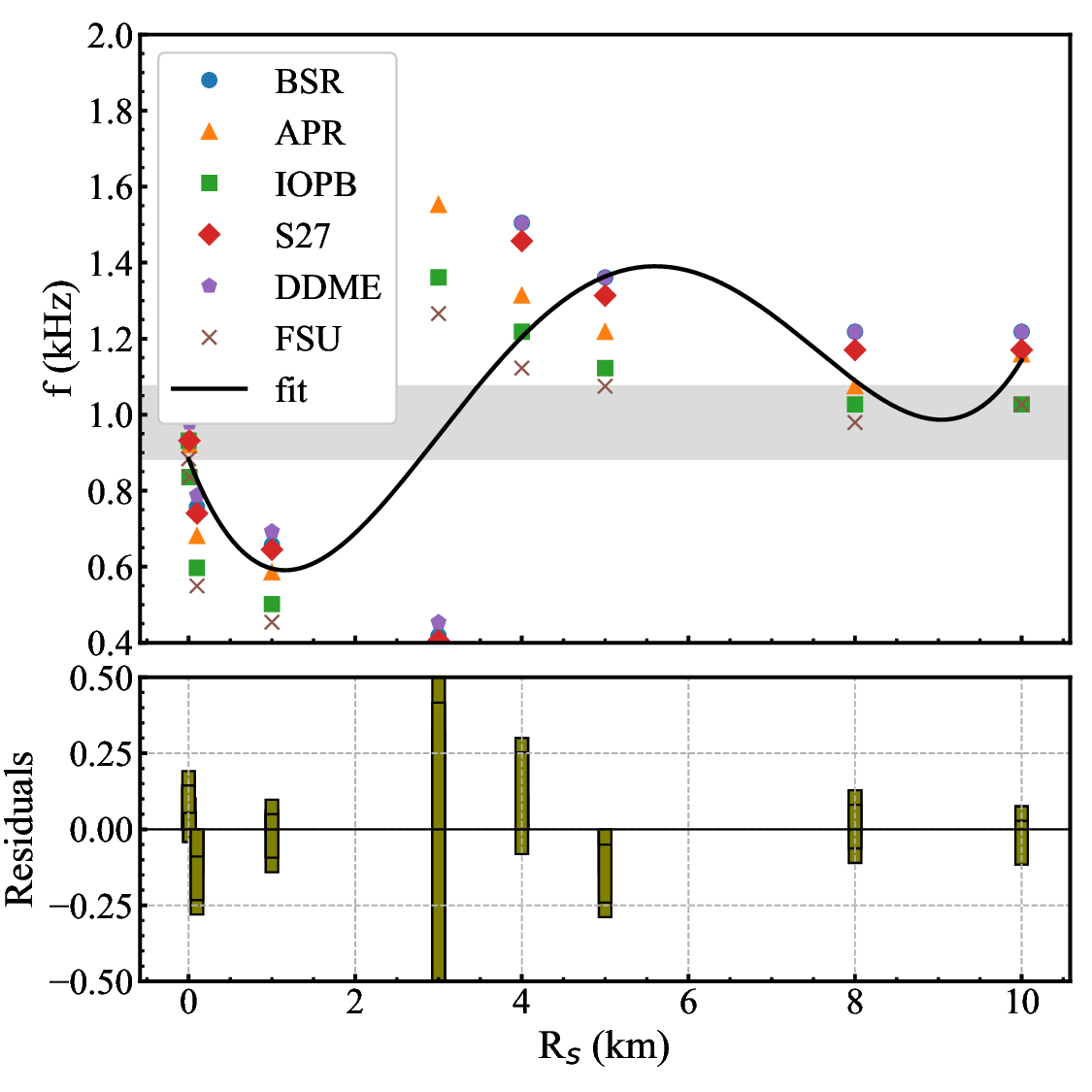}
    \caption{This figure examines how the radial f-mode frequency, denoted as \( f \), varies as a function of the shell radius \( R_s \), for different equations of state (EoS). The f-mode frequencies were extracted from the radial oscillation analysis for NS models containing a singular thin shell, introduced via a delta-type energy density at radius \( R_s \). The shaded grey band represents the frequency range corresponding to standard no-shell configurations, providing a benchmark to assess whether singular shell models produce frequencies indistinguishable from normal NSs.}
    \label{fig4}
\end{figure}

The eigenvalues $\omega^2$ is shown in the upper panel of figure \ref{fig4} which demonstrates a non-monotonic variation of \( f=\frac{\omega}{2\pi} \) with shell radius \( R_s \). For small $R_s$, the f-mode values show that the NS-shell configuration remains stable with $\omega^2 > 0$, but its value drops compared to the no-shell case, which indicates that the presence of a shell closer to the core softens the NS against radial perturbations. This is primarily because the shell drops the pressure support available at the core against compression from the upper layers, which in turn causes the fluid elements of the NS to experience a lower restoring force. This implies that for such a configuration, the fluid elements are more easily allowed to oscillate, corresponding to a lower eigenmode frequency value. As $R_s$ increases, $ f$-mode exhibits a non-linear behavior, where intermediate shell positions can either stiffen or soften the configuration depending on the interaction of the shell defect with the background pressure profile of the NS. The plot indicates that the shell acts as an effective stiffener rather than a softening agent. This is primarily due to the fact that at intermediate $R_s$ ($\approx 3-4$ km) the eigenfunction and its derivatives switch sign around such $R_s$ as can be seen in figure \ref{fig3}. This, instead of lowering the restoring force, rather reinforces it strongly. Near the stellar surface, $f$ values approach the no-shell limit, which reflects the fact that a shell located in the outermost layers has a weak influence on the overall pressure and density profile. The figure therefore makes explicit how the shell location can act as a control parameter for stability, with higher $R_s$ driving the configuration towards more stability. 

The black curve represents a fourth-order polynomial fit $ f(R_s) = a_0 + a_1 R_s + a_2 R_s^2 + a_3 R_s^3 + a_4 R_s^4 $
where \( a_0, a_1, ..., a_4 \) are fitting coefficients determined from the data across all EoS. The fitted polynomial coefficients are:
$a_4 = 2.41\times10^{-3}$, $a_3 = -5.09\times10^{-2}$, $a_2 = 3.26\times10^{-1}$, $a_1 = -5.66\times10^{-1}$, $a_0 = 8.83\times10^{-1}$. The corresponding $1-\sigma$ uncertainties are:
$\sigma_{a_4} = 5.34\times10^{-4}$, $\sigma_{a_3} = 1.08\times10^{-2}$, $\sigma_{a_2} = 7.02\times10^{-2}$, $\sigma_{a_1} = 1.47\times10^{-1}$, $\sigma_{a_0} = 5.31\times10^{-2}$ all in appropriate units.

The lower panel plots the residuals of the fit $\Delta f(R_s) = f_{\text{computed}}(R_s) - f_{\text{fit}}(R_s)  $. It can be noted that several NS-shell configurations yield frequencies that lie within the gray band (figure \ref{fig4}), i.e., within the frequency range of no-shell NSs. This region acts as an EoS-induced uncertainty in the predicted $f$-mode frequency for a NS. It is seen that the frequency shifts induced by the singular shell lie within but mostly exceeds this band, depending on the shell position. This indicates towards a degeneracy: that is, a configuration with a shell can mimic the oscillation spectrum of a no-shell NS constructed with a different EoS. This also implies that a lower-frequency configuration due to a core shell may resemble a softer EoS, while a higher-frequency configuration from an intermediate shell radius could dummy a stiffer EoS. Once the radial perturbation analysis is established, it becomes essential to understand these dynamical changes in terms of the broader context of the equilibrium of the NS as discussed in the next section.

\begin{figure}
    \centering
    \includegraphics[scale=0.46]{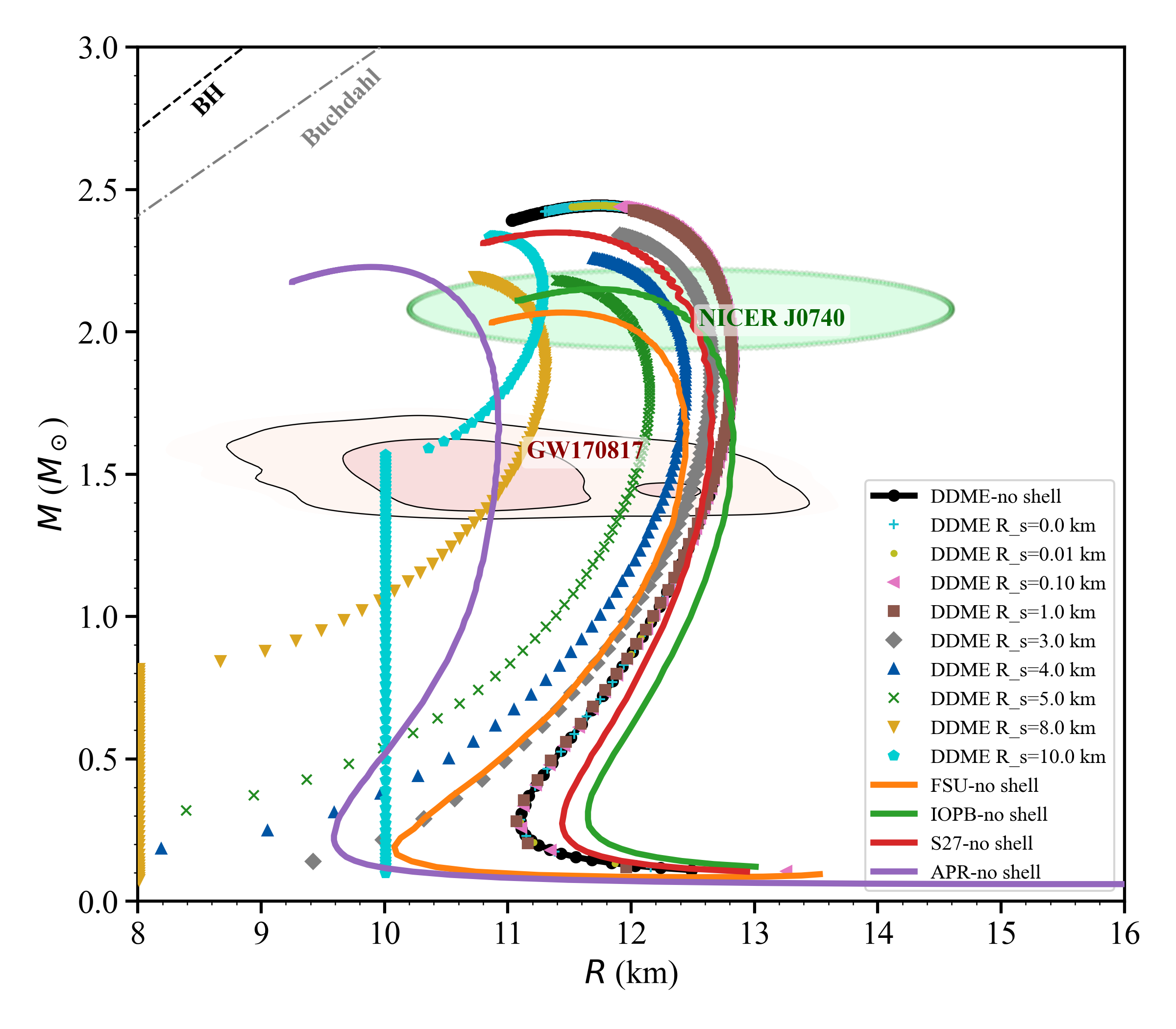}
    \caption{Mass–radius (\(M\)–\(R\)) relations for NSs constructed using the DDME EoS with and without a singular shell, along with other EoS models (FSU, IOPB, S27, APR). The curves labeled ``no shell'' represent standard NSs modeled with continuous density profiles. The discrete markers represent NSs modeled with a singular thin shell at radius \(R_s\), ranging from \(R_s = 0.01\) km to \(R_s = 10\) km. The shell configurations significantly deviate from standard profiles, particularly for large \(R_s\), exhibiting non-trivial disconnected sequences and forbidden regions in the \(M\)–\(R\) plane. The region {GW170817} \cite{Abbott2017_GW170817} and the NICER J0740 \cite{Miller_2021,Riley_2019} are the observational constraints along with theoretical limits of Buchdahl \cite{Dadhich_2020}, Schwarzschild (BH) ($R=2M$) are overlayed.}
    \label{fig5}
\end{figure}

\section{Equilibrium Structure and Mass-Radius Relations}
\subsubsection{MR Curve}
The effect of the shell on the overall stellar structure is presented in figure \ref{fig5}, where the mass–radius (MR) relation for equilibrium configurations with different shell radii $R_s$ has been displayed. For the no–shell case, the curve follows the usual pattern, where the mass at first rises with central density, which leads to a maximum, after which it decreases. The lower radius region after the maximum represents the onset of instability in the usual TOV sense. When a shell is introduced, first, the maximum mass reachable by the NS for a particular EoS is slightly shifted, reflecting the redistribution of the pressure profile. Shells placed at lower $R_s$ reduce the effective central stiffness and thus lower the peak mass, whereas shells for higher $R_s$ again could be seen having negligible influence on the maximum mass. Secondly, the radius of a given mass NS changes depending on $R_s$. Say, for  lower shells radius $R_s$, a given central density (and thus pressure) drops to a certain value, which is still significant enough to keep hold of the matter beyond the $R_s$, and almost stable branches of MR curves could be seen. However, as $R_s$ keeps on increasing, the pressure fall happens at a much higher radius, towards the surface of the NS, where the pressure is already very less. Hence, for lower central densities, it can be seen in figure \ref{fig5} that, NSs keeps on forming with radius equal to $R_s$ (prominent for $R_s=8$ km and $10$ km), and only when the central density is high enough to hold the matter beyond the $R_s$, does the MR curve produces the standard characteristic curve.

The figure, along with the equilibrium characteristics, also displays 
a set of observational and theoretical constraints in the MR plane. 
The region {GW170817} corresponds to the LIGO-Virgo 90\% credible posterior plot from the binary NS merger GW170817 \cite{Abbott2017_GW170817}. Along with this, the NICER J0740+6620 posterior plot corresponds to an accurate radius measurement for a $\sim 2\,M_\odot$ pulsar \cite{Miller_2021,Riley_2019}. In addition to this, two important theoretical limits have been overlayed, that is Buchdahl limit \cite{Dadhich_2020}, and the Schwarzschild black hole (BH) line ($R=2M$). Such characteristics broaden the possible MR band, producing configurations that could overlap with those of different EoS families. All together, these results reinforce the degeneracy highlighted in figure \ref{fig4}, that the presence of a shell can mimic the structural signature of either a softer or stiffer EoS.

\begin{figure*}
    \centering
    \includegraphics[scale=0.40]{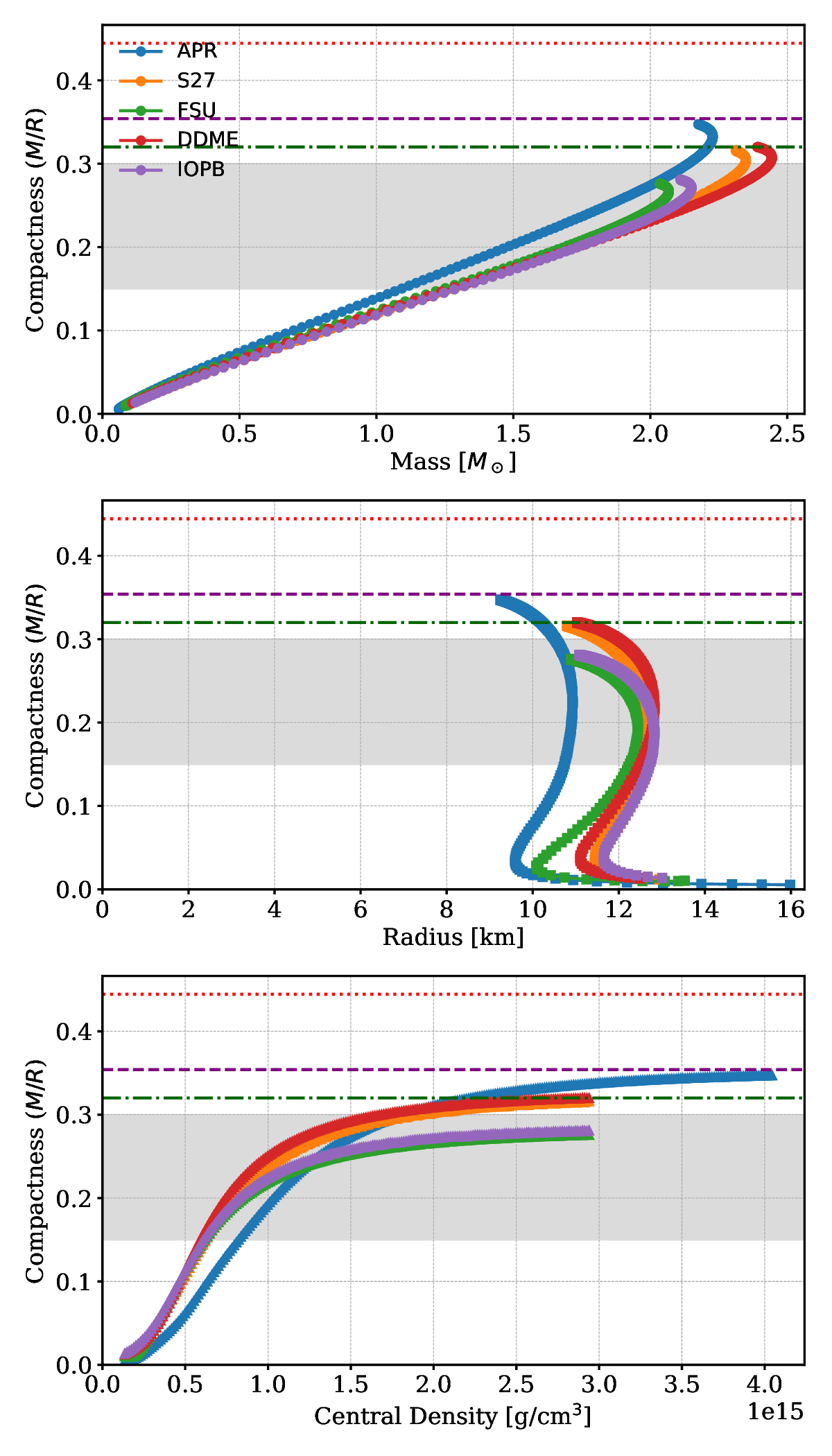}        \includegraphics[scale=0.40]{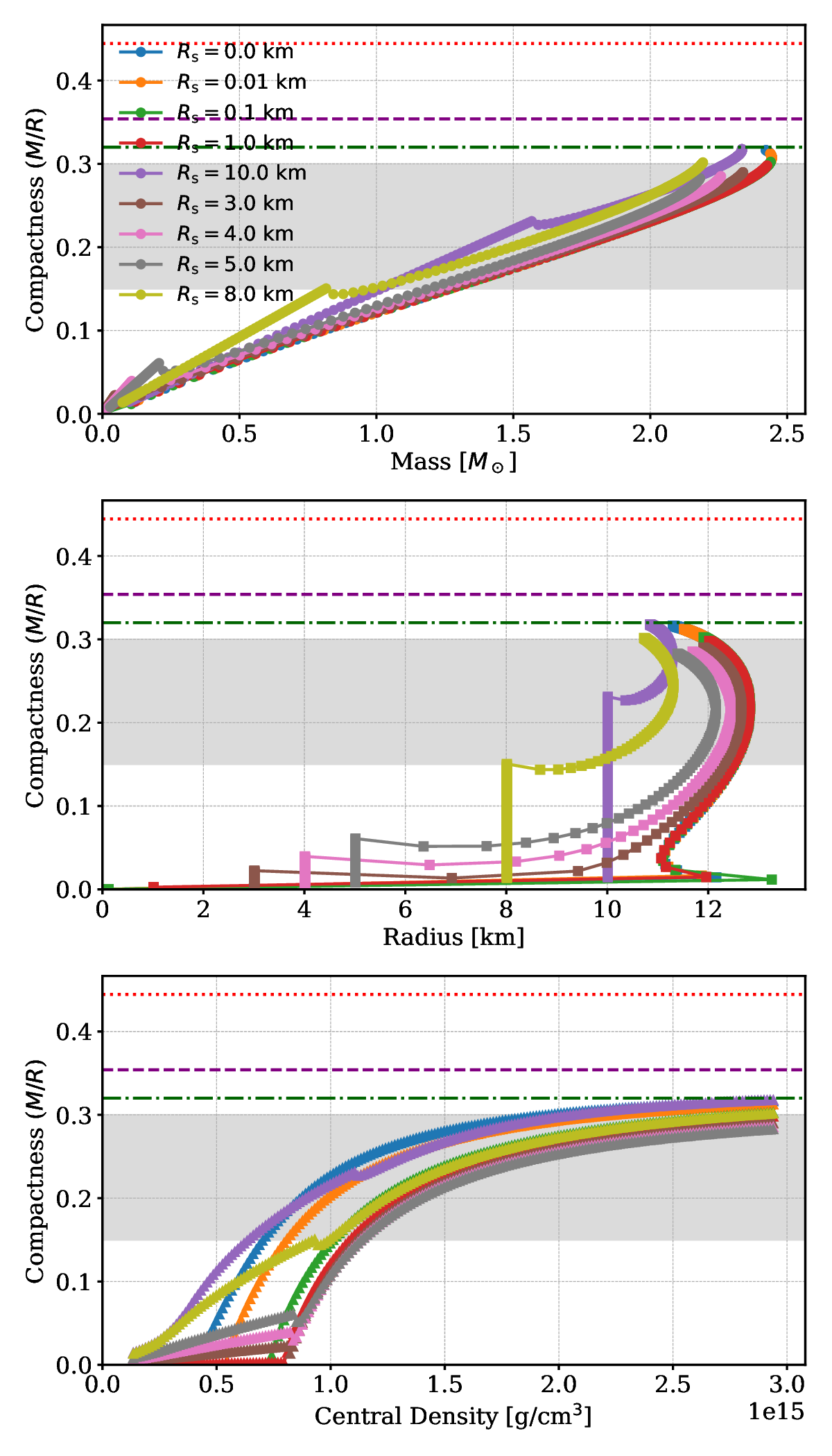}
    \caption{Compactness (\(M/R\)) variation for NS models without and with singular shell structures.
Top panels: Compactness as a function of gravitational mass \(M\). 
Middle panels: Compactness as a function of stellar radius \(R\). 
Bottom panels: Compactness as a function of central density \(\rho_c\). 
Left-hand panels show results for different EoS without shells (APR, S27, IOPB, DDME, FSU, BSR), while right-hand panels show DDME EoS results for varying shell radii \(R_s\). 
Critical compactness limits are marked: 
red dotted line (Buchdahl limit: \(M/R = 4/9\)), 
purple dashed line (Causal limit: \(M/R \approx 0.354\)), 
green dash-dotted line (Realistic EoS limit: \(M/R \approx 0.32\)), 
and grey shaded region indicating observationally realistic compactness range (\(0.15 \leq M/R \leq 0.30\)).
}
    \label{fig6}
\end{figure*}
\subsubsection{Compactness Trends and Constraints}
The influence of the shell on the stellar compactness is shown in figure \ref{fig6}, with six panels plotting $C=M/R$ as a function of gravitational mass (top left), stellar radius (middle left), and central density (bottom left), together with corresponding panels (right side) that represent the equivalent NS-shell case.  
In the no–shell scenario, for the bottom left plot, $C$ follows the expected pattern, which is generally seen in the case of an NS, that is, it increases monotonically with central density, reaches a maximum near the maximum–mass configuration, and lies in the typical range $0.1 \lesssim C \lesssim 0.3$. When a shell is introduced, it systematically shifts these relations, where shells near the core, as discussed in the previous section, make the core less stiff and hence increase the compactness for a given mass or central density, also reflecting the reduction in stellar radius. 
Intermediate–radius shells, on the other hand, act to decrease $C$, as now the NS has a larger radius even for the same enclosed mass. Shells located in the outer layers have only a minor effect, leaving the curves essentially unchanged. Various recent astrophysical and theoretical constraints have been overlapped in the plots. The horizontal line at $C=4/9$ marks the Buchdahl limit \cite{Dadhich_2020}, beyond which, no physically stable (non black-hole) solution can exceed this compactness, that is, a purely geometric upper bound. The Causal limit is $C\lesssim 0.354$ \cite{PhysRevD.95.083014}, observational constraints from NICER radius measurements, tidal deformabilities extracted from the GW170817, and pulsar mass observations, are represented by the grey band ($C_{1.4} \simeq 0.15-0.17$ and $C_{2.0}\simeq 0.24-0.26$) \cite{Miller_2021,Riley_2019,Abbott2017_GW170817}. For reference, the empirical realistic EoS limit is also mentioned ($C\lesssim 0.32$) \cite{Köppel_2019}.  

Our results show that the presence of a shell moves configurations towards or away from these bounds depending on its location. Shells near the core push the models to higher compactness, potentially closer to the causality line, while intermediate shells shift them toward lower compactness values consistent with softer EoS. Importantly, it can be seen that all models remain within the Buchdahl limit and within the observationally inferred window as discussed; however, the spread induced by the shell overlaps with the uncertainty due to the EoS. Thus, interestingly, compactness constraints alone cannot disentangle the microphysics of the EoS from the macroscopic effects of singular shells, and some form of degeneracy remains. The structural deviation from the NS-shell configurations will naturally propagate to the oscillatory and GW properties of the NS. Particularly, singular shells mimicking softer or stiffer EoS indicate that such topological defects may introduce various degeneracies in GW-based detectors.

\begin{figure}
    \centering
    \includegraphics[scale=0.45]{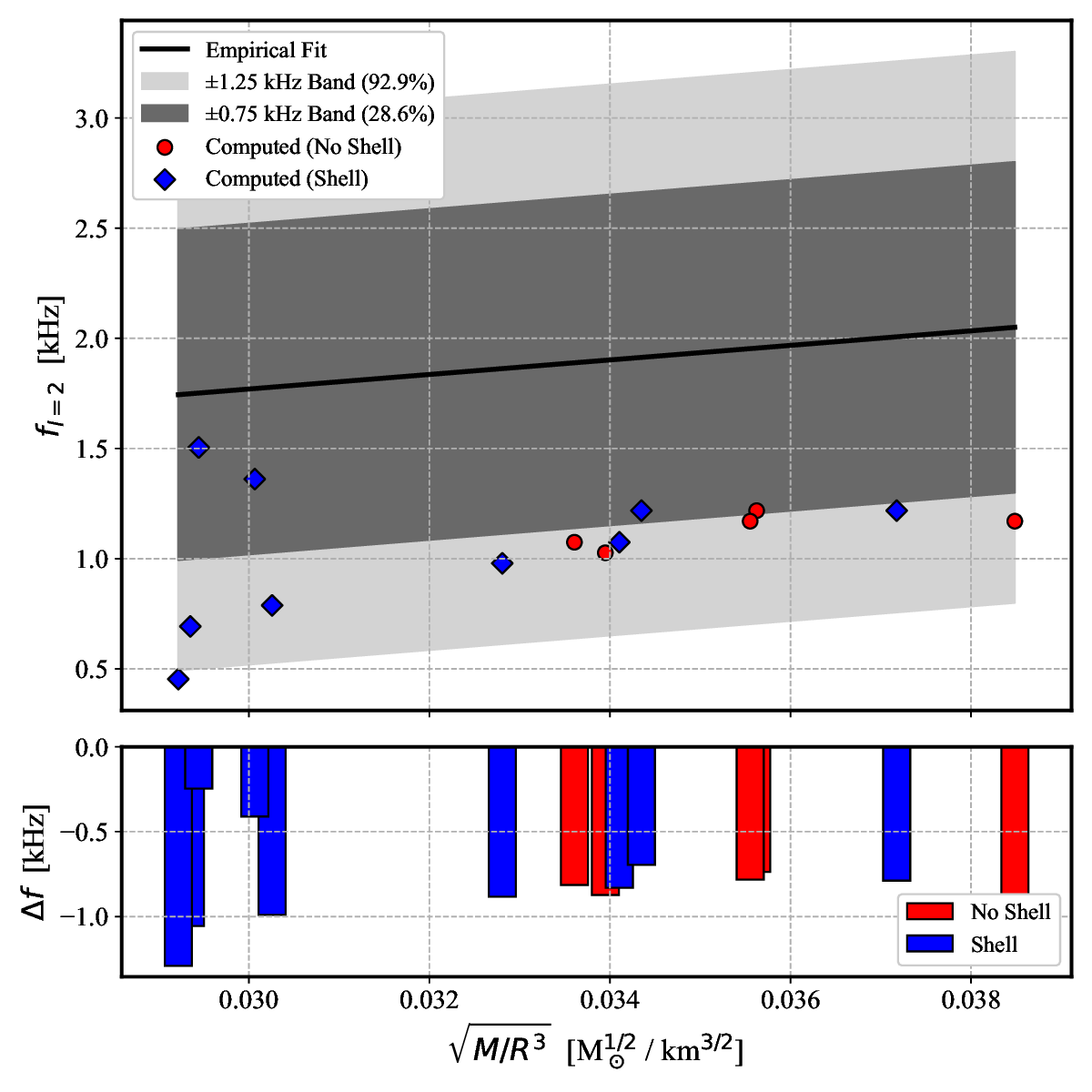}
    \caption{Comparison between computed fundamental \(f\)-mode frequencies and empirical relation.
Top panel: The computed fundamental \(f\)-mode frequencies (\(f_{l=2}\)) are plotted as a function of \(\sqrt{M/R^3}\), which represents the mean stellar density. 
Red circles denote frequencies for no-shell NS models, while blue diamonds represent models with singular shell structures. 
The empirical fit (black line) follows the relation:
$f_{l=2} = a \sqrt{\frac{M}{R^3}} + b$
where \(a\) and \(b\) are empirically determined constants. 
Two shaded confidence bands are shown: the dark grey band corresponds to a \(\pm 0.75\)~kHz interval around the fit (containing 28.6\% of the computed modes), and the lighter grey band corresponds to a broader \(\pm 1.25\)~kHz band (covering 92.9\% of the computed modes).
Bottom panel: Residuals (\(\Delta f\)) plotted as the difference between computed modes and the empirical fit. Blue bars represent shell configurations, and red bars represent no-shell models.
}
\label{fig7}
\end{figure}

\section{Extension to Non-Radial Modes and GW analysis}
\subsection{Correlating $l=0$ and $l=2$ f-Modes}
It is important to acknowledge here that the frequencies calculated throughout this work correspond to \emph{radial} oscillations ($l=0$), which preserve spherical symmetry and therefore 
have no corresponding radiation. Nevertheless, they have the same fundamental dynamical degree of freedom, in the sense that they arise from the same underlying mechanism of the pressure–gravity restoring force. To analyze how the presence of a singular shell affects the quasi-relations between different $f$-mode families, we compare our computed $l=0$ frequencies with the corresponding possible $l=2$ $f$-mode results obtained from standard relativistic stellar models in the literature. 
This is represented in figure \ref{fig7}, which shows a comparative representation of our calculated fundamental \(f\)-mode ($l=0$) oscillation frequencies, highlighting deviations from the standard empirical fit ($l=2$). The plotted empirical relation is obtained from \cite{10.1046/j.1365-8711.1998.01840.x}
\begin{equation}
    f_{l=2} = a \sqrt{\frac{M}{R^3}} + b
\end{equation}
is based on the investigated empirical dependence of the \(f\)-mode frequency ($l=2$) on the mean density of the NS . 
It can be seen that the no-shell models (red circles) closely follow the empirical fit and doesn't spread much, with most of the data points lying within the dark grey confidence band (\(\pm 0.75\)~kHz). Only two EoS models marginally deviate beyond this band. On the other hand, shell models (blue diamonds), display much deviations, particularly for configurations with intermediate shell radius or shells asymptotes towards low-density outer regions (although much further in the outer surface behaves again as no shell counterparts). Thus, many shell configurations reside outside the narrow confidence band and enter the broader \(\pm 1.25\)~kHz band, indicating a breakdown of the empirical relation in the presence of singular shell structures.
Residuals at the bottom panel of figure \ref{fig7}, $\Delta f = f_{\text{computed}} - f_{\text{empirical fit}}$ are plotted for all models, where for the no-shell case exhibit small negative residuals, indicating minor under-prediction by the empirical fit whereas, shell models (particularly intermediate shells) shows larger residuals, reflecting pronounced deviations.

The two shaded bands quantify the confidence in using the computed ($l=0$) modes. Emphasizing once again that, $l=0$ modes that we have computed don't emit any radiation, but they serve as hydrodynamic
proxies for the ($l=2$) $f$-modes that dominate gravitational-wave (GW)
emission. The shaded confidence bands represent the degree
of reliability with which the $l=0$ frequencies can be mapped onto
GW relations that typically employ the $l=2$ values. The dark grey band (\(\pm 0.75\)~kHz) where, only 28.6\% of the total computed frequencies lie, and the frequencies differ by less than $10$ \% represents higher confidence of $\gtrsim 90\%$, where as light grey band (\(\pm 1.25\)~kHz) covers 92.9\% of the computed modes, indicating deviations upto $20 \%$ implying a lower confidence of $\gtrsim 80\%$. This provides a bridge between the present analysis and the subsequent discussion of GW observables. 

\begin{figure*}
    \centering
    \includegraphics[scale=0.43]{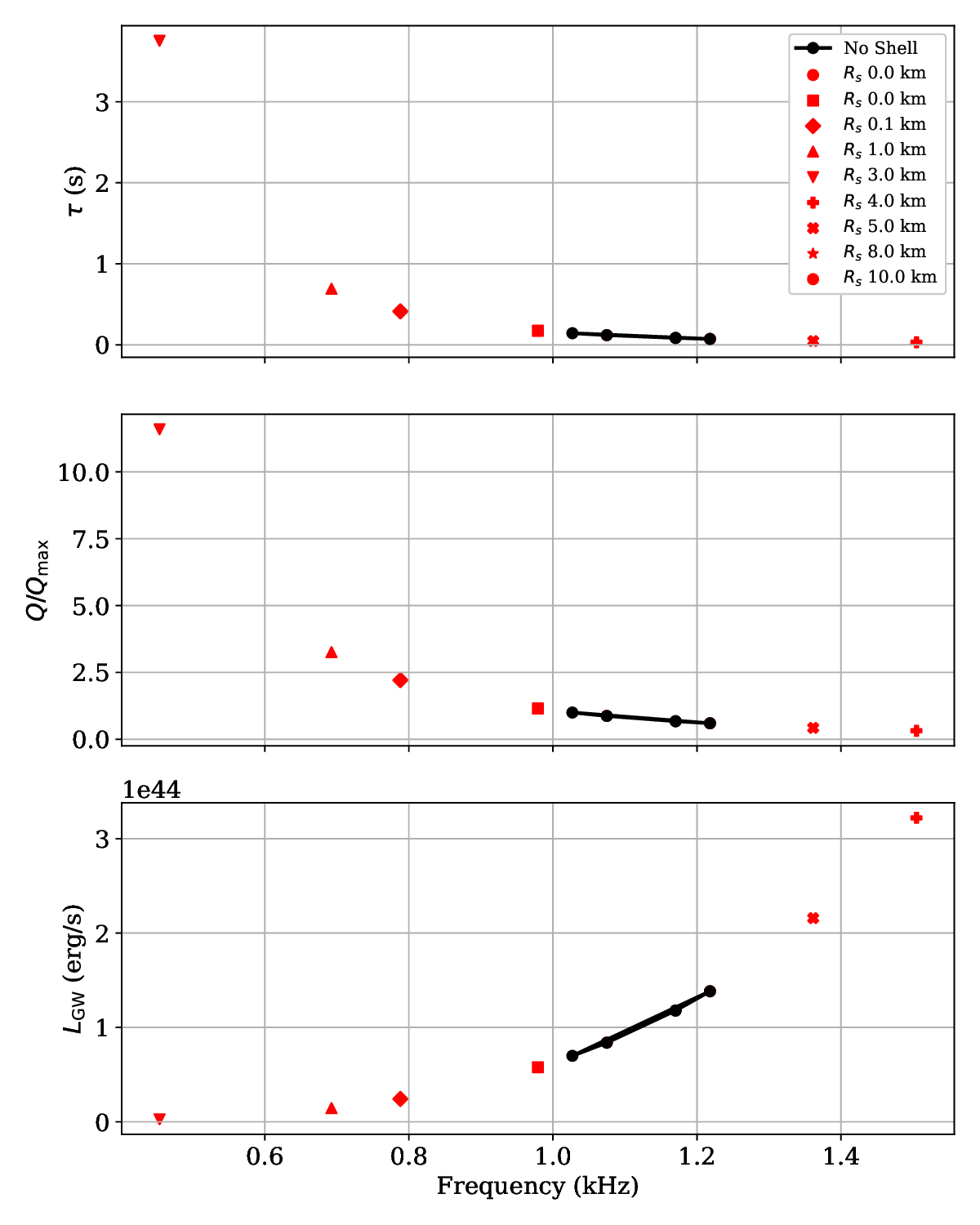}        \includegraphics[scale=0.43]{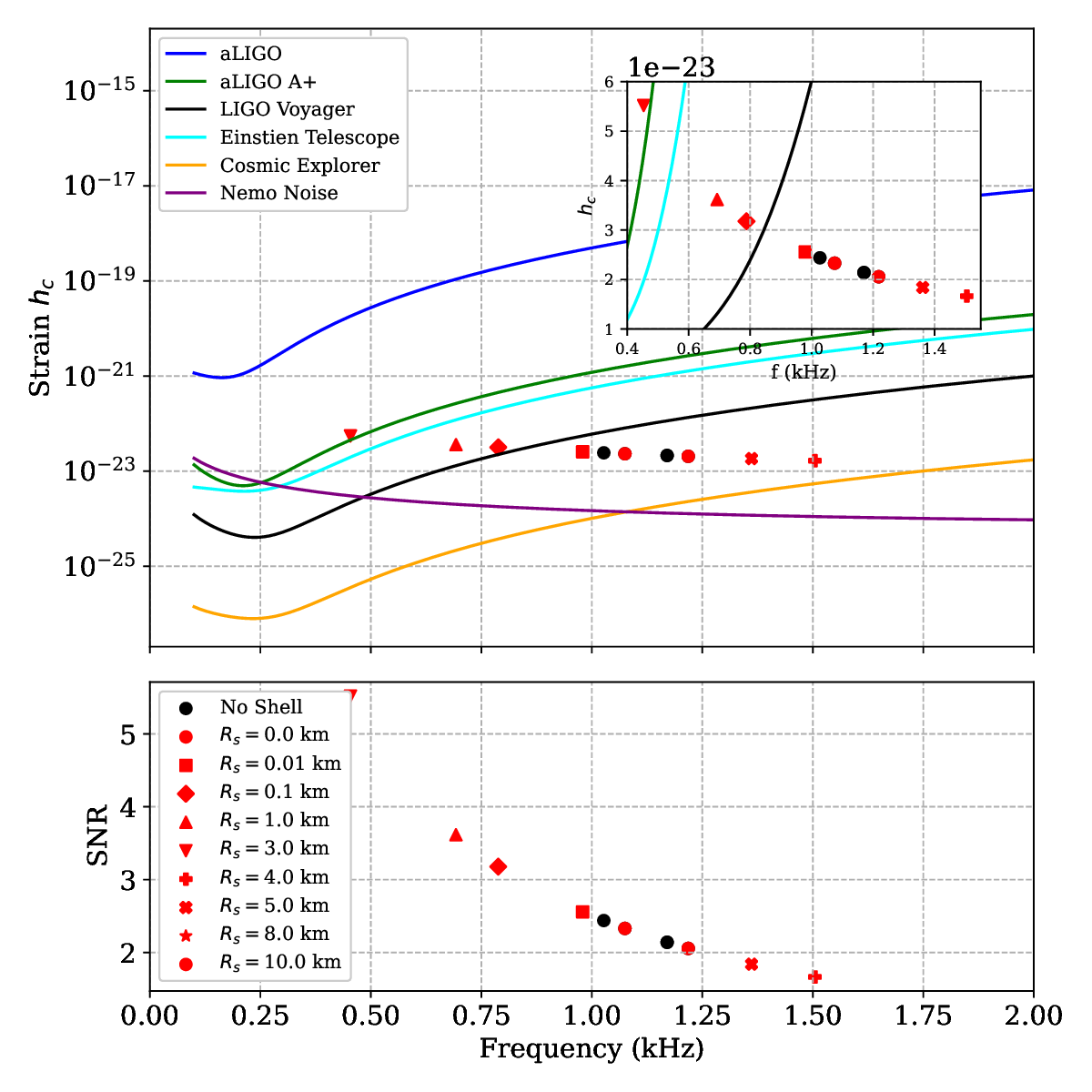}
    \caption{GW properties associated with fundamental \(f\)-mode oscillations of NSs.
Left panel (from top to bottom):
(a) Damping time \(\tau\) vs. mode frequency.
(b) Quality factor \(Q\) vs. mode frequency.
(c) GW luminosity \(L_{\text{GW}}\) vs. mode frequency.
Right panel (top):
Characteristic strain amplitude \(h_c\) vs. mode frequency, overlaid with the Advanced LIGO sensitivity curve (blue) along with other future detectors.
An inset shows a zoomed-in view around \(h_c \sim 10^{-23}\).
Right panel (bottom):
Signal-to-noise ratio (SNR) vs. mode frequency for a hypothetical detection scenario.
Different shell radii \(R_s\) configurations are distinguished using various symbols, while no-shell models are shown as black circles.
}
\label{fig8}
\end{figure*}
\subsection{GW properties of the fundamental $f$-mode}

Various GW observables associated with the computed $f$-mode for our sequence of models with and without a shell is represented in figure \ref{fig8}. The left column displays (from top to bottom) the damping time of the modes $\tau$, the quality factor $Q$, and the GW luminosity $L_{\rm GW}$. The right column (top panel) shows the characteristic strain $h_c(f)$ overlaid with the Advanced-LIGO sensitivity curve; on the other hand, (bottom panel) the signal-to-noise ratio (SNR) for a detector configuration and source distance. These quantities are computed using the approximated relations for 
weakly damped quasi-normal oscillations. The damping time could be taken to approximately scale as
\begin{equation}
\tau \propto f^{-4}    
\end{equation}
which comes from the frist principle estimates where, the GW power is $\dot{E}_{GW}=\frac{G}{5c^5}\omega^6 Q_0^2$ \cite{RevModPhys.52.299}, the quadruple amplitude $Q_0\sim M R^2\eta$ \cite{PhysRevD.108.024003}, where $\eta$ is the radial displacement amplitude of the outer layer of the NS, that give rise to the quadruple moment and is related to our eigenfunction $\zeta$. On the other hand, the mode or the oscillation energy can be scaled as $E_{osc}\sim\frac{1}{2}\omega^2M\eta^2$, where, now eliminating $\eta$ gives, $\dot{E}_{GW}\propto \omega^4E_{osc}$, implying $\tau=E_{osc}/\dot{E}_{GW}\propto f^{-4}$. This shows that modes of higher-frequency decay faster. The shell models exhibit a higher damping time relative to the no-shell cases due to their lower frequency. However, this tendency manifests itself till upto around $\approx 4.0$ km, after which the damping time dips to a minimum, even getting below the no-shell case, due to a sudden jump in the mode frequency. This tendency, however, resorts back towards the no-shell case as the shell radius moves towards the surface of the NS. For such damping times, the quality factor is defined as \cite{Kokkotas:2007zz}, 
\begin{equation}
 Q=\pi f \tau   
\end{equation}
which characterizes the number of oscillation cycles before the amplitude decays substantially. This again reflects the damping profile of the mode frequencies, although it can be seen that, $Q$ scales as $\sim f^{-2}$ in contrast to $\sim f^{-3}$. The GW luminosity is estimated from the quadrupole formula, scaling as
\begin{equation}
  L_{\rm GW} \propto f^{6} \,\eta^{2},  
\end{equation}
Since also $L_{\rm GW} \propto f^{4} E_{\rm osc}$, the luminosity can be seen increasing 
steeply with frequency, with the high frequency modes radiating significantly more power. For shell configurations, the luminosity again tracks the damping time behavior and that of the mode frequency. For Shells which are located near the stellar core (small $R_s$), as they possess lower mode frequencies, thus emit comparatively less gravitational power. As the shell radius increases, the frequencies and thus $L_{\rm GW}$ follow an increasing trend; however, as discussed earlier that around intermediate radii ($R_s\!\approx\!4$--$5$~km), the mode frequencies do undergo a sharp transition and luminosity eventually reverts toward the no-shell configuration. 

In general, the luminosity trends along with the $\tau$ and $Q$ provide the interpretation that singular shell defects can, as a matter of fact, act as amplifiers or dampers 
of the GW emission efficiency, depending on their location within the 
stellar interior.

The right column of figure \ref{fig8} presents the characteristic strain $h_c(f)$ and the corresponding signal-to-noise ratio (SNR) for the computed $f$-modes, evaluated for both the no-shell and shell configurations. These quantities provide direct observational measures of the GW detectability of the oscillations by ground-based detectors such as Advanced LIGO.

The characteristic strain, shown in the top right panel, is estimated using the standard weak-damping approximated model for f-mode oscillation \cite{PhysRevD.111.063069},
\begin{equation}
h_c(f) = \frac{1}{D}\sqrt{\frac{2 G E_{\rm osc}}{\pi^2 f^2 c^3 \tau}},
\end{equation}
where $D$ is the source distance. In this work, we have taken $D = 10~{\rm kpc}$ and $E_{\rm osc} = 10^{-8} M_{\odot} c^2$ as typical parameters for a Galactic source, which correspond to excitation energies expected from possible NS dynamics such as crustal quakes, phase transitions, or glitch-induced core oscillations etc. The resulting $h_c(f)$ curves are shown where the Advanced-LIGO sensitivity band is overlapped, to illustrate their relative detectability \cite{PhysRevD.102.062003}. 

The strain amplitude is seen to scale inversely as a function of the distance, roughly as $h_c \propto f^{-1}\tau^{-1/2}$, implying that modes with higher frequency and short lives have intrinsically weaker strain amplitudes despite their larger luminosity. 
Thus, as a result, configurations with smaller $\tau$ (such as the ones with intermediate radius shells exhibiting higher mode frequencies) produce a noticeably reduced strain amplitude compared to low-frequency, 
longer-lived modes. However, for Galactic sources, the resulting $h_c$ values still fall within or close to the detection threshold of advanced interferometers if the excitation energy exceeds $E_{\rm osc} \gtrsim 10^{-8} M_{\odot} c^2$.

We have also compared our calculated strain with respect to the targeted sensitivity of the future planned ground-based GW detectors such as Advanced-LIGO (O5 design) \cite{PhysRevD.109.103535}, its upgrade A+ \cite{PhysRevD.91.062005}, the proposed LIGO-Voyager configuration \cite{Adhikari_2020}, and the third-generation observatories Einstein Telescope (ET) \cite{Branchesi_2023}, Cosmic Explorer (CE) \cite{galaxies10040090}, and the high-frequency concept NEMO \cite{Ackley_2020}. We here have used analytic sensitivity fits that approximately reproduce the projected strain noise across the kilohertz band, so that it can be 
compared with the predicted $f$-mode strains from our models.

It is seen that the calculated f-modes approach the detection threshold of its A+ and Voyager upgrades, in particular for the Galactic sources ($D\simeq10~\mathrm{kpc}$), accompanied with excitation energies of $E_{\rm osc}\sim10^{-8}M_{\odot}c^2$. In contrast, detectors such as ET and CE display strain sensitivities nearly an order of magnitude 
better in the $1$--$3$~kHz range, which places our NS-shell models 
well within their detectable region, with respect to the no-shell models. The conceptually planned NEMO detector, which is optimized for high-frequency signals, indicates promising prospects for direct observation of such $f$-mode signals in future runs.

The lower right panel in Figure \ref{fig8} represents the corresponding signal-to-noise ratios (SNR) computed for the same detector configurations as discussed earlier. This is defined as,
\begin{equation}
{\rm SNR} = \frac{h_c(f)}{\sqrt{S_n(f)}},
\end{equation}
where $S_n(f)$ denotes the detector’s one-sided noise power spectral density. For analysis purposes, we have assumed a frequency-independent noise strain $1/\sqrt{S_n(f)}=\alpha$, where $\alpha$ is of the order of $\sim 10^{23}$ and is used solely to illustrate the relative 
change in signal strength between shell and no–shell configurations. The SNR values overall follow the trend of $h_c(f)$ and show that core-radius shells, which produce 
lower-frequency and low luminous oscillations, yield the largest SNR values. 
In contrast, intermediate shells, associated with higher oscillation 
frequencies or lower damping times, result in smaller SNRs. Overall, the 
figure \ref{fig8} demonstrates that the detectability of NS-shell $f$-modes improves significantly in future detectors, which suggests that the presence of localized internal structures could be observationally constrained through future high-sensitivity GW observations.

\begin{figure}
    \centering
    \includegraphics[scale=0.35]{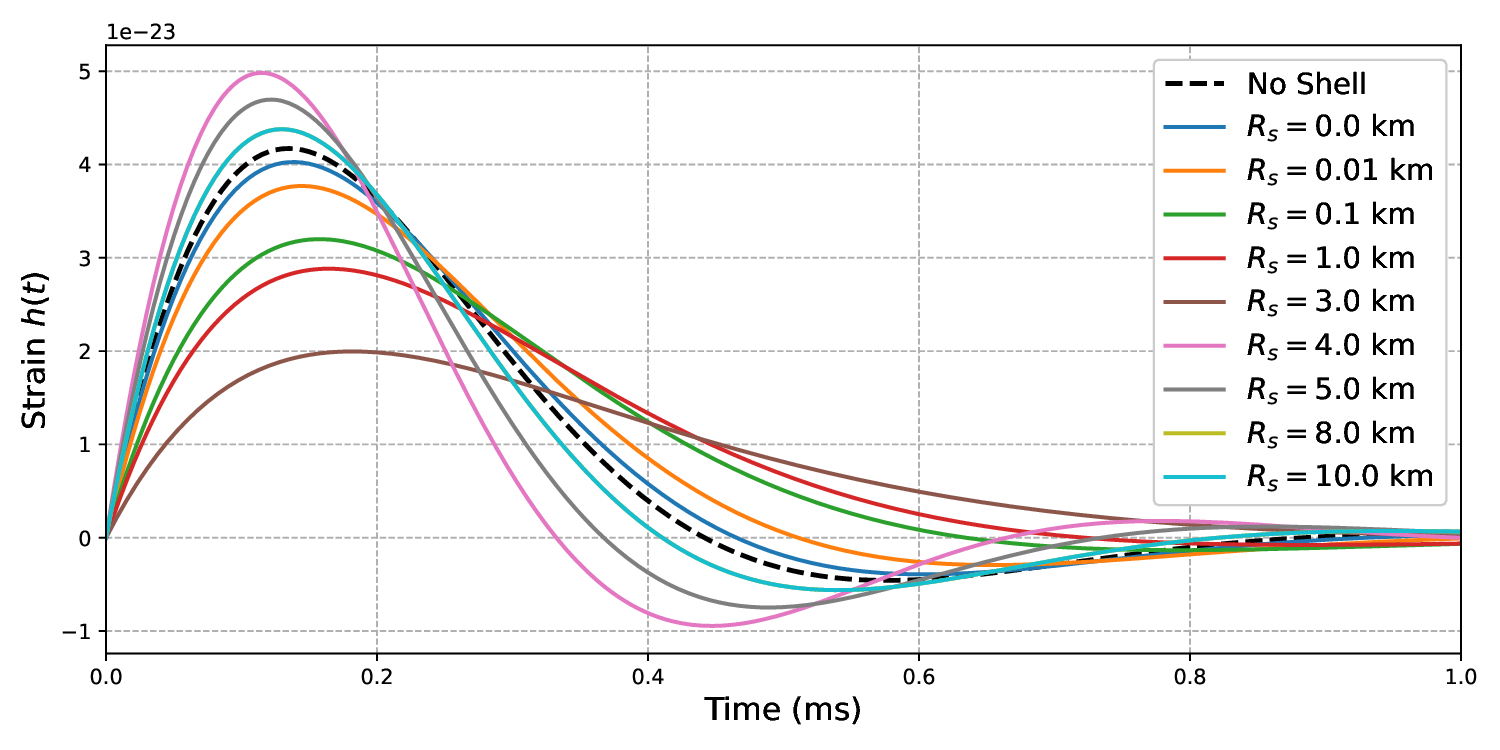}
    \caption{Time-domain GW strain signal from fundamental \(f\)-mode oscillations. The figure shows the strain amplitude \(h(t)\) as a function of time for NSs modeled both with and without a singular shell. 
The no-shell waveform (black dashed line) serves as a reference, while distinct colored curves represent different shell radius configurations (\(R_s\)). 
Shell structures significantly alter the amplitude and damping profile of the emitted GWs, with larger shell radii generally leading to earlier decay and amplitude modulation.}
    \label{fig9}
\end{figure}
\subsubsection{Waveform morphology of f-mode ringdown type signals}
The GW strain's temporal profile, which is generated by the $f$-mode 
oscillations of our NS-shell/no shell models, as is shown in figure \ref{fig9}. The standard definition of quasi-normal oscillations can be represented by a damped sinusoidal waveform,
\begin{equation}
h(t) = h_0\, e^{-t/\tau}\cos\!\left(2\pi f\,t + \phi_0\right),
\label{eq:ringdown}
\end{equation}
where $f$ and $\tau$ denote the mode frequency and damping time obtained from the previous analysis, $h_0$ is the peak amplitude, and $\phi_0$ is the initial phase. The peak strain amplitude $h_0$, as used previously, is related to the total energy radiated in GWs as $h_0 = \frac{1}{D}\sqrt{\frac{2GE_{\rm GW}}{\pi^2 f^2 c^3 \tau}}$. We have used the same estimates of the parameters as used earlier, corresponding to a typical Galactic event.

The waveforms in figure \ref{fig9} display the characteristic damped oscillatory behaviour of $f$-mode ringdown type signals. The decay time scale of the amplitude scales as $\tau$, while the oscillation frequency of the wavefront determines the spacing between successive crossings of the zero mark. The shell configurations that are located near the core 
is seen to typically produce slightly lower-frequency and longer-lived signals; on the other hand, intermediate shells are seen to yield higher-frequency oscillations accompanied by faster exponential decay. 
These features reflect the trends discussed in the previous section, where the damping time $\tau$ was shown to vary roughly as $\tau\propto f^{-4}$.

For the no-shell case, the strain decays smoothly with a typical timescale of $\tau\!\sim\!0.1$--$0.3~\mathrm{s}$, corresponding to well-established $f$-mode damping behaviour. The inclusion of a localized shell modifies both the amplitude and decay rate. Core-shell configurations show lower maximum amplitude, but slower amplitude decay, which implies slightly longer persistence, contrary to which, intermediate-radius shells, although they exhibit higher maximum amplitude, rapidly drop within a few tens of milliseconds. Now, coming to the surface shells, it  restores to the behavior of a no-shell profile, which again shows that the perturbation of the outer layers does not significantly alter the global dynamics. 

Overall, figure \ref{fig9} represents that the encoding of the internal structure of such NS-shell configurations is embedded in the time-domain waveform parametrized by both its frequency and damping envelope. The distinctive temporal signatures in principle could leave observable imprints in future high-sensitivity detectors, providing a potential probe of internal discontinuities in NSs.

\section{Conclusion}

In this work, we have developed and analyzed a class of NS models containing a localized singular mass shell, represented by delta-type density distributions at an arbitrary radius $R_s$. Such a construction effectively mimics a thin, finite-energy membrane embedded within an otherwise continuous fluid interior, thereby enabling a consistent general relativistic treatment of internal structural discontinuities. The formalism incorporates the shell directly into the Tolman–Oppenheimer–Volkoff (TOV) framework through a pressure-jump condition that emerges from the discontinuity in the stress–energy 
tensor at $r = R_s$. This approach permits the study of how delta-like mass layers influence the global equilibrium, stability, and oscillation properties of compact stars, without invoking external matching or separate spacetime patches. 

By solving the modified TOV equations with the pressure discontinuity, we 
constructed equilibrium sequences of NSs with different shell radii and 
examined the resulting mass–radius ($M$–$R$) relations. The presence of the singular shell was found to alter both the gravitational mass and the stellar radius in a nontrivial way: shells located near the core generally soften the NS, reducing its maximum mass and radius, whereas intermediate shells tend to expand the configuration for a fixed central density. The shell’s influence diminishes as it approaches the surface, where the structure gradually converges to the no-shell configuration. These effects are quantitatively reflected in the compactness–mass and compactness–radius diagrams, which also demonstrate consistency with current astrophysical constraints from pulsar timing and NICER observations.

The analysis of the radial oscillation spectrum reveals that the fundamental ($f$) mode frequencies are significantly affected by the shell position. Configurations with shells deep inside the core exhibit lower mode frequencies due to the softening of the inner density profile, while those with intermediate shells experience a sharp increase in frequency as a result of the sign reversal in the displacement eigenfunctions and their derivatives. Importantly, all computed $\omega^2$ values remain positive, confirming the stability of the equilibrium within the considered parameter range. The gray band in the frequency plots marks the maximum frequency variation across different equations of state for the no-shell cases, highlighting that the introduction of a singular shell breaks this degeneracy and introduces a measurable spread in the $f$-mode spectrum.

In terms of global stellar observables, the compactness parameter 
$C = GM/(Rc^2)$ exhibits distinct behavior across shell radii. Shells near the core increase $C$ by concentrating the mass internally, whereas the intermediate shells decrease $C$ by expanding the stellar envelope. The inclusion of astrophysical constraints, such as causality limits, rotation-supported bounds, and observational inferences from PSR~J0740+6620 and GW170817, confirms that the shell models remain physically admissible within current empirical uncertainties. 

The GW analysis showed that how these internal 
discontinuities reciprocates into observational signatures. The $f$-mode damping times, quality factors, and luminosities were computed using various scaling relations for weakly damped oscillations, with the physically motivated scaling $\tau \propto f^{-4}$. The luminosity was obtained as $L_{\rm GW} = E_{\rm GW}/\tau$, assuming an energy conversion efficiency $\epsilon \sim 10^{-3}$. Models with intermediate shells show an increase in corresponding $L_{\rm GW}$ due to their higher frequencies and shorter damping times, whereas core and surface shells display lower emission. The computed characteristic strain and SNR, evaluated for a Galactic source at $10~{\rm kpc}$ and $E_{\rm osc} = 10^{-8} M_\odot c^2$, suggest that such oscillations may fall within the sensitivity band of current interferometers if the mode excitation is energetic enough. 

Finally, the time-domain GW strain illustrates how the presence of a shell alters the temporal morphology of the emitted signal. The waveform retains its exponentially decaying sinusoidal form, but exhibits distinct envelope shapes for different shell locations. 
Core shells lead to long-lived, slowly damped oscillations with lower frequencies, while intermediate shells produce rapidly decaying high-frequency bursts. Surface shells restore a nearly symmetric decay pattern similar to that of standard NSs. These wavefront modifications demonstrate that localized internal structures imprint recognizable temporal signatures on the ringdown signal.

In summary, our study establishes that singular mass shells can modify both the static and dynamical behavior of NSs without compromising stability. Their signatures appear not only in equilibrium properties but also in the observable characteristics of GW emission. The distinctive features identified here, like the variation in decay envelope and oscillation period, suggest that future high-sensitivity GW observations could provide indirect evidence for internal density discontinuities or phase-layered structures within NS interiors. Future work will extend this framework to include rotational and magnetic effects and explore the coupling between the $\ell=0$ and radiative $\ell=2$ sectors for a complete characterization of shell-induced GW signals.

\section*{Acknowledgment}
DK would like to thank Raghunathpur College and Department of Physics, for providing the infrastructure for this work. DK would also like to acknowledge the support in form of discussions at the International Center for Theoretical Sciences (ICTS) by participating in the program - The Future of Gravitational - Wave Astronomy 2025 (code: ICTS/FGWA2025/10). KKN would like to acknowledge the Department of Atomic Energy (DAE), Govt. of India, for sponsoring the fellowship covered under the sub-project no. RIN4001-SPS (Basic research in Physical Sciences). KKN would like to thank Deeptak Biswas for meaningful suggestions. 
\bibliography{top}

@article{10.1093/mnras/stae1258,
    author = {Lieu, Richard},
    title = {The binding of cosmological structures by massless topological defects},
    journal = {Monthly Notices of the Royal Astronomical Society},
    volume = {531},
    number = {1},
    pages = {1630-1636},
    year = {2024},
    month = {05},
    issn = {0035-8711},
    doi = {10.1093/mnras/stae1258},
    url = {https://doi.org/10.1093/mnras/stae1258},
    eprint = {https://academic.oup.com/mnras/article-pdf/531/1/1630/58552931/stae1258.pdf},
}

@ARTICLE{2020PhRvC.102a5802M,
       author = {{Mondal}, C. and {Vi{\~n}as}, X. and {Centelles}, M. and {De}, J.~N.},
        title = "{Structure and composition of the inner crust of neutron stars from Gogny interactions}",
      journal = {\prc},
         year = 2020,
        month = jul,
       volume = {102},
       number = {1},
          eid = {015802},
        pages = {015802},
          doi = {10.1103/PhysRevC.102.015802},
archivePrefix = {arXiv},
       eprint = {2003.03338},
 primaryClass = {nucl-th},
       adsurl = {https://ui.adsabs.harvard.edu/abs/2020PhRvC.102a5802M}
}

@ARTICLE{1966NCimB..44....1I,
       author = {{Israel}, W.},
        title = "{Singular hypersurfaces and thin shells in general relativity}",
      journal = {Nuovo Cimento B Serie},
         year = 1966,
        month = jul,
       volume = {44},
       number = {1},
        pages = {1-14},
          doi = {10.1007/BF02710419},
       adsurl = {https://ui.adsabs.harvard.edu/abs/1966NCimB..44....1I}
}

@ARTICLE{2025PhRvD.111l4017A,
       author = {{Arrechea}, Julio and {Barcel{\'o}}, Carlos and {Garc{\'\i}a-Moreno}, Gerardo and {Polo-G{\'o}mez}, Jos{\'e}},
        title = "{Beyond Buchdahl's limit: Bilayered stars and thin-shell configurations}",
      journal = {\prd},
         year = 2025,
        month = jun,
       volume = {111},
       number = {12},
          eid = {124017},
        pages = {124017},
          doi = {10.1103/dqbd-9z5f},
archivePrefix = {arXiv},
       eprint = {2411.14018},
 primaryClass = {gr-qc},
       adsurl = {https://ui.adsabs.harvard.edu/abs/2025PhRvD.111l4017A}
}

@ARTICLE{1999IJMPD...8..549Z,
       author = {{Zloshchastiev}, Konstantin G.},
        title = "{Barotropic Thin Shells with Linear EOS as Models of Stars and Circumstellar Shells in General Relativity}",
      journal = {International Journal of Modern Physics D},
         year = 1999,
        month = jan,
       volume = {8},
       number = {4},
        pages = {549-555},
          doi = {10.1142/S0218271899000389},
archivePrefix = {arXiv},
       eprint = {gr-qc/9802041},
 primaryClass = {gr-qc},
       adsurl = {https://ui.adsabs.harvard.edu/abs/1999IJMPD...8..549Z}
}

@ARTICLE{1964RSPSA.282..303B,
       author = {{Bondi}, H.},
        title = "{Massive Spheres in General Relativity}",
      journal = {Proceedings of the Royal Society of London Series A},
         year = 1964,
        month = nov,
       volume = {282},
       number = {1390},
        pages = {303-317},
          doi = {10.1098/rspa.1964.0234},
       adsurl = {https://ui.adsabs.harvard.edu/abs/1964RSPSA.282..303B}
}

@ARTICLE{2023Univ....9..305P,
       author = {{Pereira}, Jonas P. and {Rueda}, Jorge A.},
        title = "{Matching Slowly Rotating Spacetimes Split by Dynamic Thin Shells}",
      journal = {Universe},
         year = 2023,
        month = jun,
       volume = {9},
       number = {7},
          eid = {305},
        pages = {305},
          doi = {10.3390/universe9070305},
archivePrefix = {arXiv},
       eprint = {2306.15455},
 primaryClass = {gr-qc},
       adsurl = {https://ui.adsabs.harvard.edu/abs/2023Univ....9..305P}
}

@ARTICLE{2002PhRvD..66j4020H,
       author = {{Hosotani}, Yutaka and {Nakajima}, Takayuki and {Daghigh}, Ramin G. and {Kapusta}, Joseph I.},
        title = "{Cosmic shells}",
      journal = {\prd},
         year = 2002,
        month = nov,
       volume = {66},
       number = {10},
          eid = {104020},
        pages = {104020},
          doi = {10.1103/PhysRevD.66.104020},
archivePrefix = {arXiv},
       eprint = {gr-qc/0112079},
 primaryClass = {gr-qc},
       adsurl = {https://ui.adsabs.harvard.edu/abs/2002PhRvD..66j4020H}
}

@article{PhysRevD.43.1129,
  title = {Thin shells in general relativity and cosmology: The lightlike limit},
  author = {Barrab\`es, C. and Israel, W.},
  journal = {Phys. Rev. D},
  volume = {43},
  issue = {4},
  pages = {1129--1142},
  numpages = {0},
  year = {1991},
  month = {Feb},
  publisher = {American Physical Society},
  doi = {10.1103/PhysRevD.43.1129},
  url = {https://link.aps.org/doi/10.1103/PhysRevD.43.1129}
}

@ARTICLE{1964ApJ...140..417C,
       author = {{Chandrasekhar}, S.},
        title = "{The Dynamical Instability of Gaseous Masses Approaching the Schwarzschild Limit in General Relativity.}",
      journal = {\apj},
         year = 1964,
        month = aug,
       volume = {140},
        pages = {417},
          doi = {10.1086/147938},
       adsurl = {https://ui.adsabs.harvard.edu/abs/1964ApJ...140..417C}
}

@article{Rather_2024,
doi = {10.1088/1475-7516/2024/05/130},
url = {https://doi.org/10.1088/1475-7516/2024/05/130},
year = {2024},
month = {may},
publisher = {IOP Publishing},
volume = {2024},
number = {05},
pages = {130},
author = {Rather, Ishfaq A. and Marquez, Kauan D. and Backes, Betânia C. and Panotopoulos, Grigoris and Lopes, Ilídio},
title = {Radial oscillations of hybrid stars and neutron stars including delta baryons: the effect of a slow quark phase transition},
journal = {Journal of Cosmology and Astroparticle Physics}
}

@article{PhysRevD.103.103003,
  title = {Equation of state and radial oscillations of neutron stars},
  author = {Sun, Ting-Ting and Zheng, Zi-Yue and Chen, Huan and Burgio, G. Fiorella and Schulze, Hans-Josef},
  journal = {Phys. Rev. D},
  volume = {103},
  issue = {10},
  pages = {103003},
  numpages = {10},
  year = {2021},
  month = {May},
  publisher = {American Physical Society},
  doi = {10.1103/PhysRevD.103.103003},
  url = {https://link.aps.org/doi/10.1103/PhysRevD.103.103003}
}

@ARTICLE{1966ApJ...145..505B,
       author = {{Bardeen}, James M. and {Thorne}, Kip S. and {Meltzer}, David W.},
        title = "{A Catalogue of Methods for Studying the Normal Modes of Radial Pulsation of General-Relativistic Stellar Models}",
      journal = {\apj},
         year = 1966,
        month = aug,
       volume = {145},
        pages = {505},
          doi = {10.1086/148791},
       adsurl = {https://ui.adsabs.harvard.edu/abs/1966ApJ...145..505B}
}

@article{10.1093/mnras/stab050,
    author = {Bora, Jyatsnasree and Dev Goswami, Umananda},
    title = {Radial oscillations and gravitational wave echoes of strange stars for various equations of state},
    journal = {Monthly Notices of the Royal Astronomical Society},
    volume = {502},
    number = {2},
    pages = {1557-1568},
    year = {2021},
    month = {01},
    issn = {0035-8711},
    doi = {10.1093/mnras/stab050},
    url = {https://doi.org/10.1093/mnras/stab050},
    eprint = {https://academic.oup.com/mnras/article-pdf/502/2/1557/36193663/stab050.pdf},
}

@article{10.1111/j.1365-2966.2007.11625.x,
    author = {Glampedakis, Kostas and Andersson, Nils},
    title = {Lagrangian perturbation theory for rotating magnetic stars},
    journal = {Monthly Notices of the Royal Astronomical Society},
    volume = {377},
    number = {2},
    pages = {630-644},
    year = {2007},
    month = {04},
    issn = {0035-8711},
    doi = {10.1111/j.1365-2966.2007.11625.x},
    url = {https://doi.org/10.1111/j.1365-2966.2007.11625.x},
    eprint = {https://academic.oup.com/mnras/article-pdf/377/2/630/3485456/mnras0377-0630.pdf},
}

@article{PhysRevC.76.045801,
  title = {Nonrotating and rotating neutron stars in the extended field theoretical model},
  author = {Dhiman, Shashi K. and Kumar, Raj and Agrawal, B. K.},
  journal = {Phys. Rev. C},
  volume = {76},
  issue = {4},
  pages = {045801},
  numpages = {13},
  year = {2007},
  month = {Oct},
  publisher = {American Physical Society},
  doi = {10.1103/PhysRevC.76.045801},
  url = {https://link.aps.org/doi/10.1103/PhysRevC.76.045801}
}

@article{PhysRevC.66.064302,
  title = {Relativistic random-phase approximation with density-dependent meson-nucleon couplings},
  author = {Nik\ifmmode \check{s}\else \v{s}\fi{}i\ifmmode \acute{c}\else \'{c}\fi{}, T. and Vretenar, D. and Ring, P.},
  journal = {Phys. Rev. C},
  volume = {66},
  issue = {6},
  pages = {064302},
  numpages = {13},
  year = {2002},
  month = {Dec},
  publisher = {American Physical Society},
  doi = {10.1103/PhysRevC.66.064302},
  url = {https://link.aps.org/doi/10.1103/PhysRevC.66.064302}
}

@article{PhysRevC.97.045806,
  title = {New relativistic effective interaction for finite nuclei, infinite nuclear matter, and neutron stars},
  author = {Kumar, Bharat and Patra, S. K. and Agrawal, B. K.},
  journal = {Phys. Rev. C},
  volume = {97},
  issue = {4},
  pages = {045806},
  numpages = {16},
  year = {2018},
  month = {Apr},
  publisher = {American Physical Society},
  doi = {10.1103/PhysRevC.97.045806},
  url = {https://link.aps.org/doi/10.1103/PhysRevC.97.045806}
}

@article{PhysRevC.66.055803,
  title = {Constraining URCA cooling of neutron stars from the neutron radius of ${}^{208}\mathrm{Pb}$},
  author = {Horowitz, C. J. and Piekarewicz, J.},
  journal = {Phys. Rev. C},
  volume = {66},
  issue = {5},
  pages = {055803},
  numpages = {8},
  year = {2002},
  month = {Nov},
  publisher = {American Physical Society},
  doi = {10.1103/PhysRevC.66.055803},
  url = {https://link.aps.org/doi/10.1103/PhysRevC.66.055803}
}

@ARTICLE{1971ApJ...170..299B,
       author = {{Baym}, Gordon and {Pethick}, Christopher and {Sutherland}, Peter},
        title = "{The Ground State of Matter at High Densities: Equation of State and Stellar Models}",
      journal = {\apj},
         year = 1971,
        month = dec,
       volume = {170},
        pages = {299},
          doi = {10.1086/151216},
       adsurl = {https://ui.adsabs.harvard.edu/abs/1971ApJ...170..299B}
}

@article{PhysRevC.58.1804,
  title = {Equation of state of nucleon matter and neutron star structure},
  author = {Akmal, A. and Pandharipande, V. R. and Ravenhall, D. G.},
  journal = {Phys. Rev. C},
  volume = {58},
  issue = {3},
  pages = {1804--1828},
  numpages = {0},
  year = {1998},
  month = {Sep},
  publisher = {American Physical Society},
  doi = {10.1103/PhysRevC.58.1804},
  url = {https://link.aps.org/doi/10.1103/PhysRevC.58.1804}
}

@Article{Marmorini2024,
author={Marmorini, Giacomo
and Yasui, Shigehiro
and Nitta, Muneto},
title={Pulsar glitches from quantum vortex networks},
journal={Scientific Reports},
year={2024},
month={Apr},
day={03},
volume={14},
number={1},
pages={7857},
issn={2045-2322},
doi={10.1038/s41598-024-56383-w},
url={https://doi.org/10.1038/s41598-024-56383-w}
}

@article{PhysRevResearch.4.013026,
  title = {Rotating self-gravitating Bose-Einstein condensates with a crust: A model for pulsar glitches},
  author = {Verma, Akhilesh Kumar and Pandit, Rahul and Brachet, Marc E.},
  journal = {Phys. Rev. Res.},
  volume = {4},
  issue = {1},
  pages = {013026},
  numpages = {6},
  year = {2022},
  month = {Jan},
  publisher = {American Physical Society},
  doi = {10.1103/PhysRevResearch.4.013026},
  url = {https://link.aps.org/doi/10.1103/PhysRevResearch.4.013026}
}

@article{10.1093/mnras/stae2087,
    author = {Chatterjee, Sagnik and Nath, Kamal Krishna and Mallick, Ritam},
    title = {Deciphering accretion-driven starquakes in recycled millisecond pulsars using gravitational waves},
    journal = {Monthly Notices of the Royal Astronomical Society},
    volume = {534},
    number = {1},
    pages = {97-106},
    year = {2024},
    month = {09},
    issn = {0035-8711},
    doi = {10.1093/mnras/stae2087},
    url = {https://doi.org/10.1093/mnras/stae2087},
    eprint = {https://academic.oup.com/mnras/article-pdf/534/1/97/59115410/stae2087.pdf},
}

@article{PhysRevLett.133.241201,
  title = {Actinide-Boosting $r$ Process in Black-Hole--Neutron-Star Merger Ejecta},
  author = {Wanajo, Shinya and Fujibayashi, Sho and Hayashi, Kota and Kiuchi, Kenta and Sekiguchi, Yuichiro and Shibata, Masaru},
  journal = {Phys. Rev. Lett.},
  volume = {133},
  issue = {24},
  pages = {241201},
  numpages = {6},
  year = {2024},
  month = {Dec},
  publisher = {American Physical Society},
  doi = {10.1103/PhysRevLett.133.241201},
  url = {https://link.aps.org/doi/10.1103/PhysRevLett.133.241201}
}

@article{Combi_2023,
doi = {10.3847/1538-4357/acac29},
url = {https://doi.org/10.3847/1538-4357/acac29},
year = {2023},
month = {feb},
publisher = {The American Astronomical Society},
volume = {944},
number = {1},
pages = {28},
author = {Combi, Luciano and Siegel, Daniel M.},
title = {GRMHD Simulations of Neutron-star Mergers with Weak Interactions: r-process Nucleosynthesis and Electromagnetic Signatures of Dynamical Ejecta},
journal = {The Astrophysical Journal}
}

@article{Dadhich_2020,
doi = {10.1088/1475-7516/2020/04/035},
url = {https://doi.org/10.1088/1475-7516/2020/04/035},
year = {2020},
month = {apr},
publisher = {},
volume = {2020},
number = {04},
pages = {035},
author = {Dadhich, Naresh},
title = {Buchdahl compactness limit and gravitational field energy},
journal = {Journal of Cosmology and Astroparticle Physics}
}

@article{PhysRevD.95.083014,
  title = {Upper limit set by causality on the tidal deformability of a neutron star},
  author = {Van Oeveren, Eric D. and Friedman, John L.},
  journal = {Phys. Rev. D},
  volume = {95},
  issue = {8},
  pages = {083014},
  numpages = {12},
  year = {2017},
  month = {Apr},
  publisher = {American Physical Society},
  doi = {10.1103/PhysRevD.95.083014},
  url = {https://link.aps.org/doi/10.1103/PhysRevD.95.083014}
}

@article{Miller_2021,
doi = {10.3847/2041-8213/ac089b},
url = {https://doi.org/10.3847/2041-8213/ac089b},
year = {2021},
month = {sep},
publisher = {The American Astronomical Society},
volume = {918},
number = {2},
pages = {L28},
author = {Miller, M. C. and Lamb, F. K. and Dittmann, A. J. and Bogdanov, S. and Arzoumanian, Z. and Gendreau, K. C. and Guillot, S. and Ho, W. C. G. and Lattimer, J. M. and Loewenstein, M. and Morsink, S. M. and Ray, P. S. and Wolff, M. T. and Baker, C. L. and Cazeau, T. and Manthripragada, S. and Markwardt, C. B. and Okajima, T. and Pollard, S. and Cognard, I. and Cromartie, H. T. and Fonseca, E. and Guillemot, L. and Kerr, M. and Parthasarathy, A. and Pennucci, T. T. and Ransom, S. and Stairs, I.},
title = {The Radius of PSR J0740+6620 from NICER and XMM-Newton Data},
journal = {The Astrophysical Journal Letters}
}

@article{Riley_2019,
doi = {10.3847/2041-8213/ab481c},
url = {https://doi.org/10.3847/2041-8213/ab481c},
year = {2019},
month = {dec},
publisher = {The American Astronomical Society},
volume = {887},
number = {1},
pages = {L21},
author = {Riley, T. E. and Watts, A. L. and Bogdanov, S. and Ray, P. S. and Ludlam, R. M. and Guillot, S. and Arzoumanian, Z. and Baker, C. L. and Bilous, A. V. and Chakrabarty, D. and Gendreau, K. C. and Harding, A. K. and Ho, W. C. G. and Lattimer, J. M. and Morsink, S. M. and Strohmayer, T. E.},
title = {A NICER View of PSR J0030+0451: Millisecond Pulsar Parameter Estimation},
journal = {The Astrophysical Journal Letters}
}

@article{Abbott2017_GW170817,
  author  = {B.~P. Abbott and R. Abbott and T.~D. Abbott and et~al.},
  title   = {GW170817: Observation of Gravitational Waves from a Binary Neutron Star Inspiral},
  journal = {Phys. Rev. Lett.},
  volume  = {119},
  number  = {16},
  pages   = {161101},
  year    = {2017},
  month   = {Oct},
  doi     = {10.1103/PhysRevLett.119.161101},
  url     = {https://link.aps.org/doi/10.1103/PhysRevLett.119.161101},
  note    = {LIGO Scientific Collaboration and Virgo Collaboration}
}

@article{Köppel_2019,
doi = {10.3847/2041-8213/ab0210},
url = {https://doi.org/10.3847/2041-8213/ab0210},
year = {2019},
month = {feb},
publisher = {The American Astronomical Society},
volume = {872},
number = {1},
pages = {L16},
author = {Köppel, Sven and Bovard, Luke and Rezzolla, Luciano},
title = {A General-relativistic Determination of the Threshold Mass to Prompt Collapse in Binary Neutron Star Mergers},
journal = {The Astrophysical Journal Letters}
}

@article{10.1046/j.1365-8711.1998.01840.x,
    author = {Andersson, Nils and Kokkotas, Kostas D.},
    title = {Towards gravitational wave asteroseismology},
    journal = {Monthly Notices of the Royal Astronomical Society},
    volume = {299},
    number = {4},
    pages = {1059-1068},
    year = {1998},
    month = {10},
    issn = {0035-8711},
    doi = {10.1046/j.1365-8711.1998.01840.x},
    url = {https://doi.org/10.1046/j.1365-8711.1998.01840.x},
    eprint = {https://academic.oup.com/mnras/article-pdf/299/4/1059/3869494/299-4-1059.pdf},
}

@article{RevModPhys.52.299,
  title = {Multipole expansions of gravitational radiation},
  author = {Thorne, Kip S.},
  journal = {Rev. Mod. Phys.},
  volume = {52},
  issue = {2},
  pages = {299--339},
  numpages = {0},
  year = {1980},
  month = {Apr},
  publisher = {American Physical Society},
  doi = {10.1103/RevModPhys.52.299},
  url = {https://link.aps.org/doi/10.1103/RevModPhys.52.299}
}

@article{PhysRevD.108.024003,
  title = {Scalar mode quadrupole radiation from astronomical sources in $F(R)$ modified gravity},
  author = {Inagaki, Tomohiro and Taniguchi, Masahiko},
  journal = {Phys. Rev. D},
  volume = {108},
  issue = {2},
  pages = {024003},
  numpages = {11},
  year = {2023},
  month = {Jul},
  publisher = {American Physical Society},
  doi = {10.1103/PhysRevD.108.024003},
  url = {https://link.aps.org/doi/10.1103/PhysRevD.108.024003}
}

@article{Kokkotas:2007zz,
    author = "Kokkotas, Kostas D.",
    editor = "Praszalowicz, Michal and Kutschera, Marek and Malec, Edward",
    title = "{Gravitational waves}",
    journal = "Acta Phys. Polon. B",
    volume = "38",
    pages = "3891--3923",
    year = "2007"
}

@article{PhysRevD.111.063069,
  title = {$f$-mode oscillations of hybrid stars with pasta construction},
  author = {Zheng, Zi-Yue and Sun, Ting-Ting and Wei, Jin-Biao and Chen, Huan and Zheng, Xiao-Ping and Burgio, G. F. and Schulze, H.-J.},
  journal = {Phys. Rev. D},
  volume = {111},
  issue = {6},
  pages = {063069},
  numpages = {16},
  year = {2025},
  month = {Mar},
  publisher = {American Physical Society},
  doi = {10.1103/PhysRevD.111.063069},
  url = {https://link.aps.org/doi/10.1103/PhysRevD.111.063069}
}

@article{PhysRevD.72.044016,
  title = {Energy conditions and junction conditions},
  author = {Marolf, Donald and Yaida, Sho},
  journal = {Phys. Rev. D},
  volume = {72},
  issue = {4},
  pages = {044016},
  numpages = {9},
  year = {2005},
  month = {Aug},
  publisher = {American Physical Society},
  doi = {10.1103/PhysRevD.72.044016},
  url = {https://link.aps.org/doi/10.1103/PhysRevD.72.044016}
}

@Article{universe8050250,
AUTHOR = {Chu, Chong-Sun and Tan, Hai-Siong},
TITLE = {Generalized Darmois–Israel Junction Conditions},
JOURNAL = {Universe},
VOLUME = {8},
YEAR = {2022},
NUMBER = {5},
ARTICLE-NUMBER = {250},
URL = {https://www.mdpi.com/2218-1997/8/5/250},
ISSN = {2218-1997},
DOI = {10.3390/universe8050250}
}

@article{10.1046/j.1365-8711.2003.06057.x,
    author = {Miniutti, G. and Pons, J. A. and Berti, E. and Gualtieri, L. and Ferrari, V.},
    title = {Non-radial oscillation modes as a probe of density discontinuities in neutron stars},
    journal = {Monthly Notices of the Royal Astronomical Society},
    volume = {338},
    number = {2},
    pages = {389-400},
    year = {2003},
    month = {01},
    issn = {0035-8711},
    doi = {10.1046/j.1365-8711.2003.06057.x},
    url = {https://doi.org/10.1046/j.1365-8711.2003.06057.x},
    eprint = {https://academic.oup.com/mnras/article-pdf/338/2/389/2900982/338-2-389.pdf},
}

@article{Piro_2005,
doi = {10.1086/426682},
url = {https://doi.org/10.1086/426682},
year = {2005},
month = {feb},
publisher = {},
volume = {619},
number = {2},
pages = {1054},
author = {Piro, Anthony L. and Bildsten, Lars},
title = {Neutron Star Crustal Interface Waves},
journal = {The Astrophysical Journal}
}

@article{Verma_2025,
doi = {10.3847/1538-4357/ade9a2},
url = {https://doi.org/10.3847/1538-4357/ade9a2},
year = {2025},
month = {jul},
publisher = {The American Astronomical Society},
volume = {988},
number = {2},
pages = {258},
author = {Verma, Anshuman and Saha, Asim Kumar and Malik, Tuhin and Mallick, Ritam},
title = {Probing the Internal Structure of Neutron Stars: A Comparative Analysis of Three Different Classes of Equations of State},
journal = {The Astrophysical Journal}
}

@article{Gorda_2023,
doi = {10.3847/1538-4357/aceefb},
url = {https://doi.org/10.3847/1538-4357/aceefb},
year = {2023},
month = {sep},
publisher = {The American Astronomical Society},
volume = {955},
number = {2},
pages = {100},
author = {Gorda, T. and Hebeler, K. and Kurkela, A. and Schwenk, A. and Vuorinen, A.},
title = {Constraints on Strong Phase Transitions in Neutron Stars},
journal = {The Astrophysical Journal}
}

@article{PhysRevC.107.025801,
  title = {Investigating signatures of phase transitions in neutron-star cores},
  author = {Somasundaram, R. and Tews, I. and Margueron, J.},
  journal = {Phys. Rev. C},
  volume = {107},
  issue = {2},
  pages = {025801},
  numpages = {7},
  year = {2023},
  month = {Feb},
  publisher = {American Physical Society},
  doi = {10.1103/PhysRevC.107.025801},
  url = {https://link.aps.org/doi/10.1103/PhysRevC.107.025801}
}

@article{Prasad_2018,
doi = {10.3847/1538-4357/aabf3b},
url = {https://doi.org/10.3847/1538-4357/aabf3b},
year = {2018},
month = {may},
publisher = {The American Astronomical Society},
volume = {859},
number = {1},
pages = {57},
author = {Prasad, R. and Mallick, Ritam},
title = {Dynamical Phase Transition in Neutron Stars},
journal = {The Astrophysical Journal}
}

@article{https://doi.org/10.1002/asna.70017,
author = {Kuzur, Debojoti},
title = {Impact of Topological Structures on Neutron Star Rotation and Their Observational Significance},
journal = {Astronomische Nachrichten},
volume = {346},
number = {9},
pages = {e70017},
doi = {https://doi.org/10.1002/asna.70017},
url = {https://onlinelibrary.wiley.com/doi/abs/10.1002/asna.70017},
year = {2025}
}

@article{PhysRev.55.374,
  title = {On Massive Neutron Cores},
  author = {Oppenheimer, J. R. and Volkoff, G. M.},
  journal = {Phys. Rev.},
  volume = {55},
  issue = {4},
  pages = {374--381},
  numpages = {0},
  year = {1939},
  month = {Feb},
  publisher = {American Physical Society},
  doi = {10.1103/PhysRev.55.374},
  url = {https://link.aps.org/doi/10.1103/PhysRev.55.374}
}

@article{PhysRev.55.364,
  title = {Static Solutions of Einstein's Field Equations for Spheres of Fluid},
  author = {Tolman, Richard C.},
  journal = {Phys. Rev.},
  volume = {55},
  issue = {4},
  pages = {364--373},
  numpages = {0},
  year = {1939},
  month = {Feb},
  publisher = {American Physical Society},
  doi = {10.1103/PhysRev.55.364},
  url = {https://link.aps.org/doi/10.1103/PhysRev.55.364}
}

@article{PhysRevD.102.062003,
  title = {Sensitivity and performance of the Advanced LIGO detectors in the third observing run},
  author = {Buikema, A. et al},
  journal = {Phys. Rev. D},
  volume = {102},
  issue = {6},
  pages = {062003},
  numpages = {27},
  year = {2020},
  month = {Sep},
  publisher = {American Physical Society},
  doi = {10.1103/PhysRevD.102.062003},
  url = {https://link.aps.org/doi/10.1103/PhysRevD.102.062003}
}

@article{PhysRevD.109.103535,
  title = {Unbiased estimation of gravitational-wave anisotropies from noisy data},
  author = {Kouvatsos, Nikolaos and Jenkins, Alexander C. and Renzini, Arianna I. and Romano, Joseph D. and Sakellariadou, Mairi},
  journal = {Phys. Rev. D},
  volume = {109},
  issue = {10},
  pages = {103535},
  numpages = {12},
  year = {2024},
  month = {May},
  publisher = {American Physical Society},
  doi = {10.1103/PhysRevD.109.103535},
  url = {https://link.aps.org/doi/10.1103/PhysRevD.109.103535}
}

@article{PhysRevD.91.062005,
  title = {Prospects for doubling the range of Advanced LIGO},
  author = {Miller, John and Barsotti, Lisa and Vitale, Salvatore and Fritschel, Peter and Evans, Matthew and Sigg, Daniel},
  journal = {Phys. Rev. D},
  volume = {91},
  issue = {6},
  pages = {062005},
  numpages = {6},
  year = {2015},
  month = {Mar},
  publisher = {American Physical Society},
  doi = {10.1103/PhysRevD.91.062005},
  url = {https://link.aps.org/doi/10.1103/PhysRevD.91.062005}
}

@article{Adhikari_2020,
doi = {10.1088/1361-6382/ab9143},
url = {https://doi.org/10.1088/1361-6382/ab9143},
year = {2020},
month = {jul},
publisher = {IOP Publishing},
volume = {37},
number = {16},
pages = {165003},
author = {Adhikari et al},
title = {A cryogenic silicon interferometer for gravitational-wave detection},
journal = {Classical and Quantum Gravity}
}

@article{Branchesi_2023,
doi = {10.1088/1475-7516/2023/07/068},
url = {https://doi.org/10.1088/1475-7516/2023/07/068},
year = {2023},
month = {jul},
publisher = {IOP Publishing},
volume = {2023},
number = {07},
pages = {068},
author = {Branchesi et al},
title = {Science with the Einstein Telescope: a comparison of different designs},
journal = {Journal of Cosmology and Astroparticle Physics}
}

@Article{galaxies10040090,
AUTHOR = {Hall , Evan D.},
TITLE = {Cosmic Explorer: A Next-Generation Ground-Based Gravitational-Wave Observatory},
JOURNAL = {Galaxies},
VOLUME = {10},
YEAR = {2022},
NUMBER = {4},
ARTICLE-NUMBER = {90},
URL = {https://www.mdpi.com/2075-4434/10/4/90},
ISSN = {2075-4434},
DOI = {10.3390/galaxies10040090}
}

@article{Ackley_2020,
title={Neutron Star Extreme Matter Observatory: A kilohertz-band gravitational-wave detector in the global network}, volume={37}, DOI={10.1017/pasa.2020.39}, journal={Publications of the Astronomical Society of Australia}, author={Ackley, K. and Adya, V. B. and Agrawal, P. and Altin, P. and Ashton, G. and Bailes, M. and Baltinas, E. and Barbuio, A. and Beniwal, D. and Blair, C. and et al.}, year={2020}, pages={e047}}

\end{document}